\documentclass{aa}

\usepackage[varg]{txfonts}

\usepackage{graphicx}
\usepackage{natbib}
\usepackage{tikz}
\usepackage{here}
\usepackage{bm}
\usepackage{url}
\usepackage{verbatim}
\usepackage{hyperref}

\hypersetup{
    colorlinks=true,
    linkcolor=blue,
    filecolor=blue,      
    urlcolor=blue,
    pdfpagemode=FullScreen,
    citecolor = blue,
    }

\begin{document}

\title{A $p<0.0001$ detection of CMB cooling in galactic halos and its possible relation to dark matter.}

\author{Frode K. Hansen\inst{1},  Diego Garcia Lambas\inst{2,3,4}, Heliana E. Luparello\inst{2}, Facundo Toscano\inst{2} \and Luis A. Pereyra\inst{2}}

\institute{Institute of Theoretical Astrophysics, University of Oslo, PO Box 1029 Blindern, 0315 Oslo, Norway, \email{frodekh@astro.uio.no}
\and
Instituto de Astronomía Teórica y Experimental (IATE), CONICET-UNC, Córdoba, Argentina
\and
Observatorio Astronómico de Córdoba, UNC, Córdoba, Argentina
\and
Comisión Nacional de Actividades Espaciales (CONAE), Córdoba, Argentina}


\abstract
    {
      We confirm at the $5.7\sigma$ level previous studies reporting Cosmic Microwave Background (CMB) temperatures being significantly lower around nearby spiral galaxies than expected in the $\Lambda$CDM model. Results from our earlier work was disputed in a recent paper, but in that paper, areas far beyond the galactic halos were included in the analysis while the neighbourhood of the galaxies where the main signal is seen was disregarded. Here we limit the study to pixels well within the galactic halos, focus on galaxies in dense cosmic filaments and improve on signal-to-noise compared to previous studies. The average CMB temperature in discs around these galaxies is always much lower in Planck data than in any of the 10.000 Planck-like CMB simulations. Even when correcting for the look-elsewhere-effect, the detection is still at the $3-4\sigma$ level. We further show that the largest scales ($\ell<16$) of the Planck CMB fluctuations are more correlated with the distribution of nearby galaxies than $99.99\%$ of simulated CMB maps. We argue that the existence of a new CMB foreground cannot be ignored and a physical interaction mechanism, possibly involving dark matter, as well as linked to intergalactic magnetic fields, should be sought.
}

\authorrunning{Hansen, F. K. et al.} 
\titlerunning{A $p<0.0001$ detection of CMB cooling in galactic halos and its possible relation to dark matter.}


\maketitle

\section{Introduction}
\label{sect:intro}

In a recent paper, \cite{luparello} (in the following, L2023), we found that the Cosmic Microwave Background (CMB) temperature measured on discs around nearby spiral galaxies with redshift $z<0.015$ is significantly lower than around randomly positioned points on the sphere. The radial temperature profile around these galaxies is seen to remain low out to distances of at least 1-2 Mpc from the galactic centre, far beyond the visible galaxy. No physical mechanism for the lowering of the CMB temperature has been suggested, but given the large angular extension of the signal from the galactic centre, an explanation related to dark matter is a possibility.

In \cite{addison} (in the following, A2024), the results of L2023 were reproduced but the significance was disputed. A2024 correctly points out that error bars were underestimated in L2023 as CMB simulations were not used to calibrate the uncertainty. However, A2024 still confirmed a significant deviation from simulations when considering the area $<0.1^\circ$ around galaxies, but disregarded this part of the data. Instead a $\chi^2$ test was performed where the area from $0.1^\circ$ up to $20^\circ$ from the centres of the galaxies were used. This includes areas far outside the galactic halos and even far outside of groups and filaments were galaxies reside. If a physical mechanism associated with the galaxies is cooling CMB photons, we would expect to see the peak of the effect close to the galaxy and at least within the extended halo of each galaxy. Including an area much larger than this region in the analysis, substantially dilutes the significance. Our aim her is to focus on this inner part of the temperature profile around the galaxies.

In \cite{hansen2023} (in the following, H2023), we presented further evidence for the presence of this foreground. In particular, we found a very significant correlation between the positions of nearby galaxies and cold spots in the CMB over a large range of angular scales. None of 1000 simulated Planck CMB maps showed a similarly strong correlation with galaxy positions. We furthermore showed that such a galactic imprint can explain the presence of a large number of the so-called statistical anomalies in the CMB (see \cite{iands2018} for a review). In \cite{parameters}, we showed that the contamination from this unknown foreground does not significantly alter the Planck best fit cosmological parameters, but that it may offer an explanation for the anomalously strong variation of cosmological parameters over the sky \citep{fg}.

We followed up in \cite{garcialambas2023} (in the following, GL2024) by a study of the non-Gaussian feature known as the CMB Cold Spot \citep{vielva}, observed in the southern galactic hemisphere. We found that the only large group of spiral galaxies within a distance of 100 Mpc resides in the position of this cold spot. A simple modelling of the unknown foreground, using a negative temperature profile around galaxies in this area, rendered a structure with the same position and similar morphology as the CMB Cold Spot, giving further support to the possibility of a new CMB foreground component. Finally, in \cite{santander} (in the following, C2024), a $99.3\%$ correlation between a model of the local dark matter distribution and the CMB is found. They further show that the signal is highly consistent over Planck frequencies from $44$GHz to $217$GHz and therefore seems to follow a blackbody spectrum closely.

In summary, there are four different studies showing a correlation between galaxies/local matter distribution and CMB temperature: the strongly negative mean profiles around galaxies of L2023; the correlation between the galaxy distribution and the cold spots in the CMB, from large to small scales, of H2023; the nearby Eridanus group of large spiral galaxies positioned at the location of the CMB Cold Spot in GL2024; and the correlation between the local dark matter distribution and the CMB in C2024. It was claimed in A2024 that the significance reported in L2023 is overestimated. Here we discuss some weaknesses in the methodology of both L2023 and A2024, and follow up with a more thorough study on the significance of the detection and the statistical methodology. We will also correct for the look-elsewhere-effect, allowing the free parameters defining the galaxy sample to vary.

We present the CMB data, galaxy catalogues and samples used in section \ref{sect:data}. The statistical methodology and the different ways to obtain profiles are described in section \ref{sect:method} and the results are finally presented in section \ref{sect:results}.

\begin{figure}[htbp]
\includegraphics[width=\linewidth]{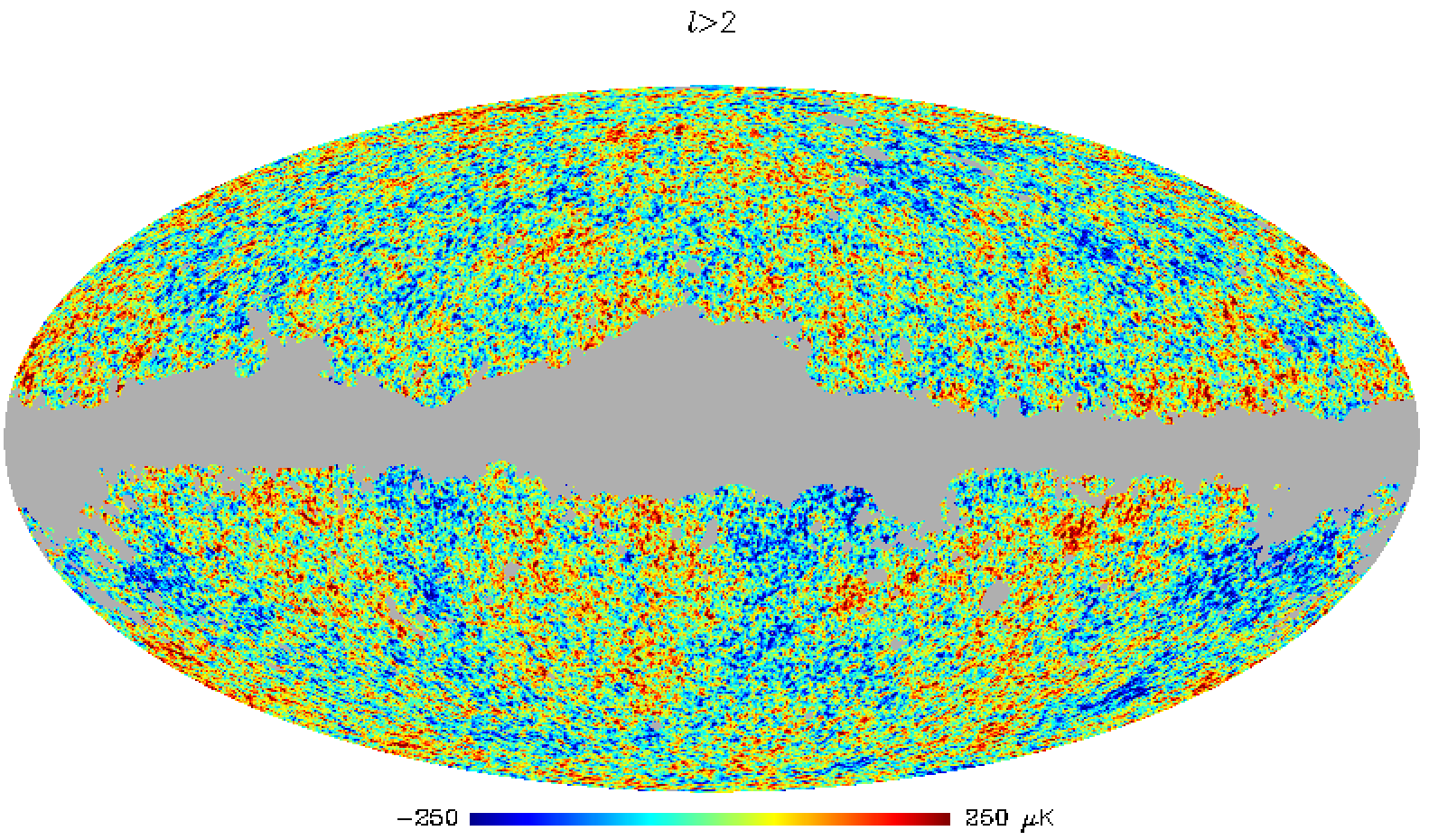}
\includegraphics[width=\linewidth]{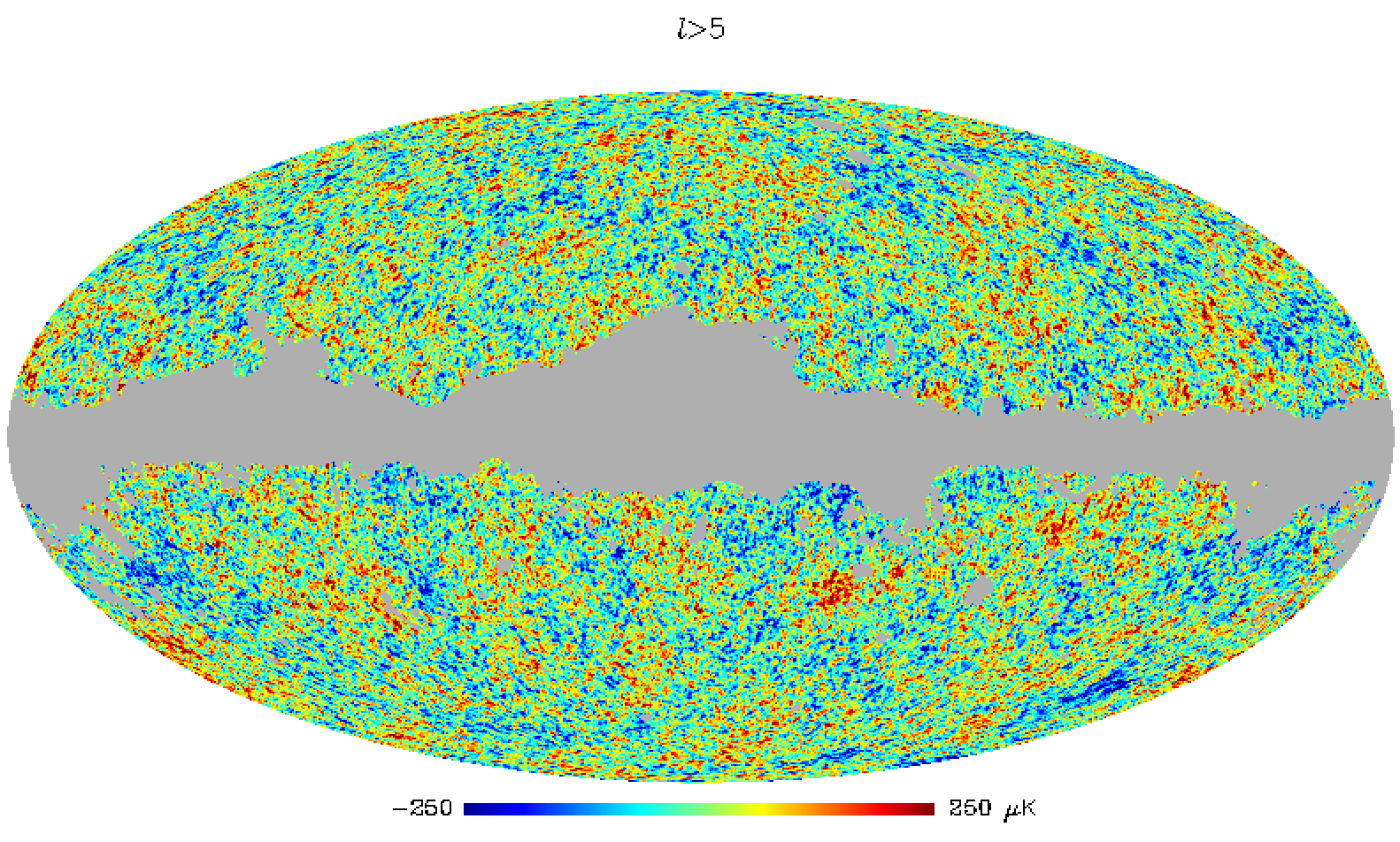}
\includegraphics[width=\linewidth]{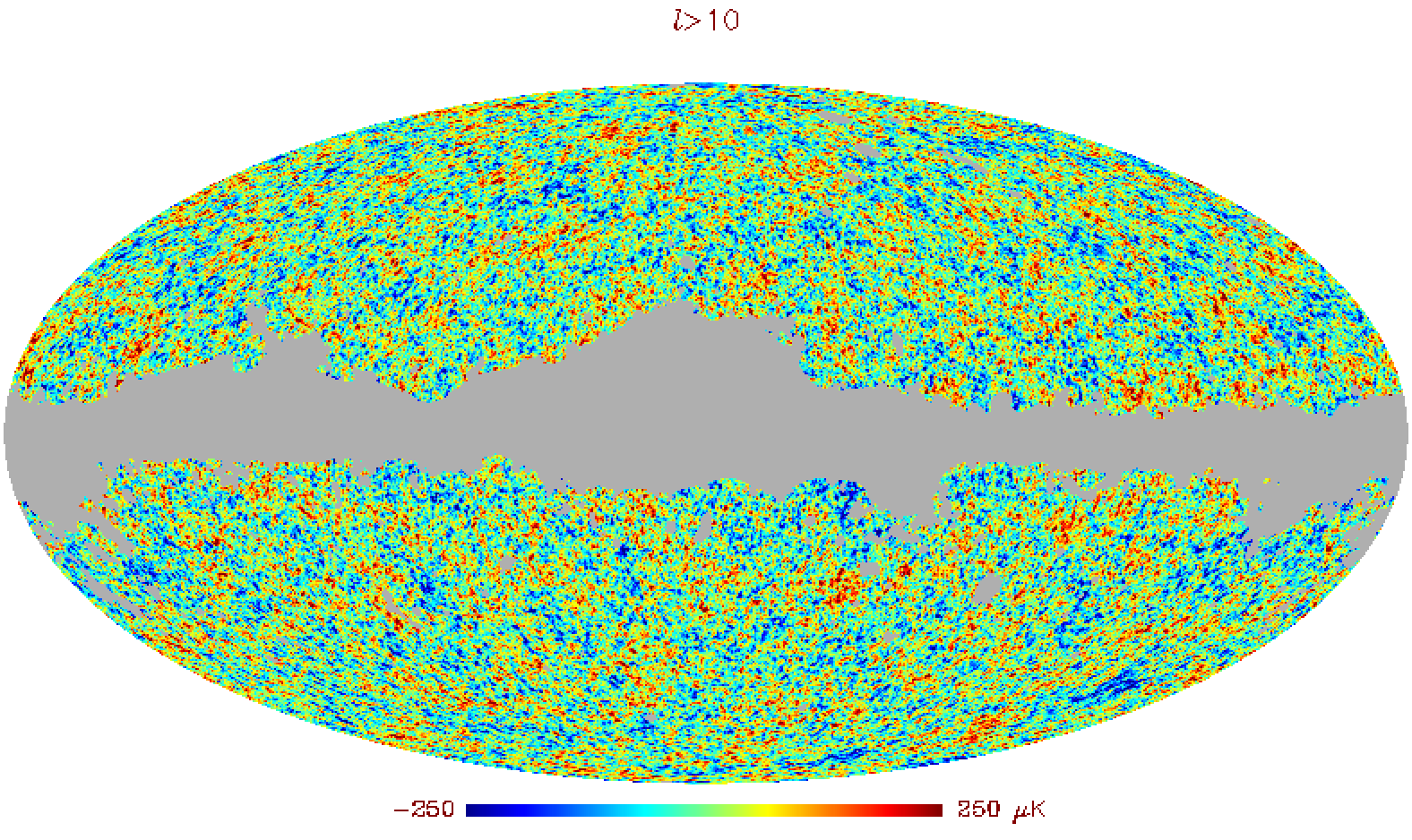}
\caption{The SMICA Planck map with $\ell\leq2$ removed (upper plot), with $\ell\leq5$ removed (middle plot) and $\ell\leq10$ removed (lower plot). Note that the small quadrupole of the actual CMB sky makes the difference between the original CMB map and the quadrupole subtracted map invisible by eye. For this reason, the full SMICA map is not shown. \label{fig:lremoved}} 
\end{figure}

\section{Data}
\label{sect:data}

In this paper, the 2MASS Redshift Survey (2MRS) \footnote{http://tdc-www.harvard.edu/2mrs/} redshift catalogue \citep{2mrs} was used to obtain the properties of nearby galaxies, including galaxy group identifications and group masses \citep{2mrsgroups}. This flux limited redshift catalogue comprises galaxy positions, redshifts, K-band magnitudes and morphological types complete up to $m_k=11.0$. By construction, this sample homogeneously covers the two galactic hemispheres with galactic latitude $|b|>10$.

We have used the Discrete Persistent Structures Extractor (DisPerSE) \footnote{http://www2.iap.fr/users/sousbie/web/html/index888d.html?archive}  \citep{Sousbie2011_a, Sousbie2011_b} filament finder to identify galactic filaments in the 2MRS catalogue, in the redshift range between $[0.004, 0.02]$. For this aim, we have considered a homogeneous sample of galaxies with radius $r > 8.5\,\, \mathrm{Kpc}$ within a deeper region, $z<0.05$, in order to avoid boundary issues. This galaxy sample is used to provide a suitable K-band galaxy density field as input to the DisPerSE code to derive the filamentary structure of the survey. We adopt a conservative persistence parameter of $4\sigma$, in order to avoid spurious structure detections, following different works in simulated and observational data \citep{Malavasi2020a, Malavasi2020b, Duckworth2020a, Duckworth2020b, GalarragaEspinosa2024}.

We acknowledge the fact that filaments have different linear mass densities of dark matter, gas and stars \citep{GalarragaEspinosa2024}. This is an important parameter for characterizing different environments which has impact on the presence of the foreground signal. We have calculated the linear K-band luminosity density within a cylindrical volume of radius $4\,\, \mathrm{Mpc}$ centred on the spine of each filament. For each filament, the total luminosity is calculated by adding together the luminosities of all galaxies located within the specified cylindrical region. In cases where a galaxy is associated with multiple filaments, its luminosity is included only in the filament that is closest to it. Finally, the total luminosity is normalised by dividing by the corresponding length of the filament to obtain the linear K-band luminosity density. 

We use the foreground cleaned CMB maps from the Planck \footnote{http://pla.esac.esa.int/pla} Public Release 3 (PR3) \citep{pr3} as well as the recent Planck Public Release 4 (PR4) \citep{pr4}. We test our results on CMB maps cleaned with all 4 official foreground methods described in \cite{compsep2018}. For SEVEM, 600 simulated maps created with the PR4 pipeline was used to calibrate uncertainties and significances. For the other three methods there were not sufficient simulations available for PR4 and we therefore used PR3 data and simulations in these cases. For NILC and Commander, there are 1000 PR3 simulations available, for SMICA an extended set of 10.000 simulations were made during the 2018 Planck release. Our main results are therefore based on SMICA to improve measures of significances, but consistency tests of all results are performed with the other methods. In particular, we always test results with the SEVEM PR4 pipeline as these maps are expected to have smaller uncertainties and foreground residuals. We also used the separate frequency cleaned maps from the SEVEM method, consisting in cleaned CMB maps at the frequencies $70$\;GHz, $100$\;GHz, $143$\;GHz and $217$\;GHz. In addition, we use the WMAP frequency cleaned maps \footnote{https://lambda.gsfc.nasa.gov/product/wmap/dr5/m\_products.html} from the 9 year release \citep{wmap9} for the frequencies $41$\;GHz, $61$\;GHz and $94$\;GHz.

For masking possible foreground residuals, we use the Planck common mask created for PR3 \citep{compsep2018} as no new mask was made available from PR4. The mask covers a large part of the galactic plane as well as resolved extragalactic point sources, leaving $\sim 78\%$ of the sky available for analysis. Note that the galactic mask to a large extent overlap with the areas also not covered with 2MASS.

Finally, for some tests we use the simplified model map of the new foreground described in H2023. As described in more detail there, this map was created by assigning spiral galaxies in the redshift range $z=[0.004,0.02]$ of type Sb, Sc and Sd of the 2MRS catalogue a linear temperature profile where the depths of the profile had a quadratic dependency on size and the angular extent of the profile had a linear dependency on the local galactic density. The properties of this model was partially based on observed properties of the foreground. This model reproduces to some extent part of the foreground signal in certain patches of the sky.

\begin{figure}[htbp]
  \includegraphics[width=0.975\linewidth]{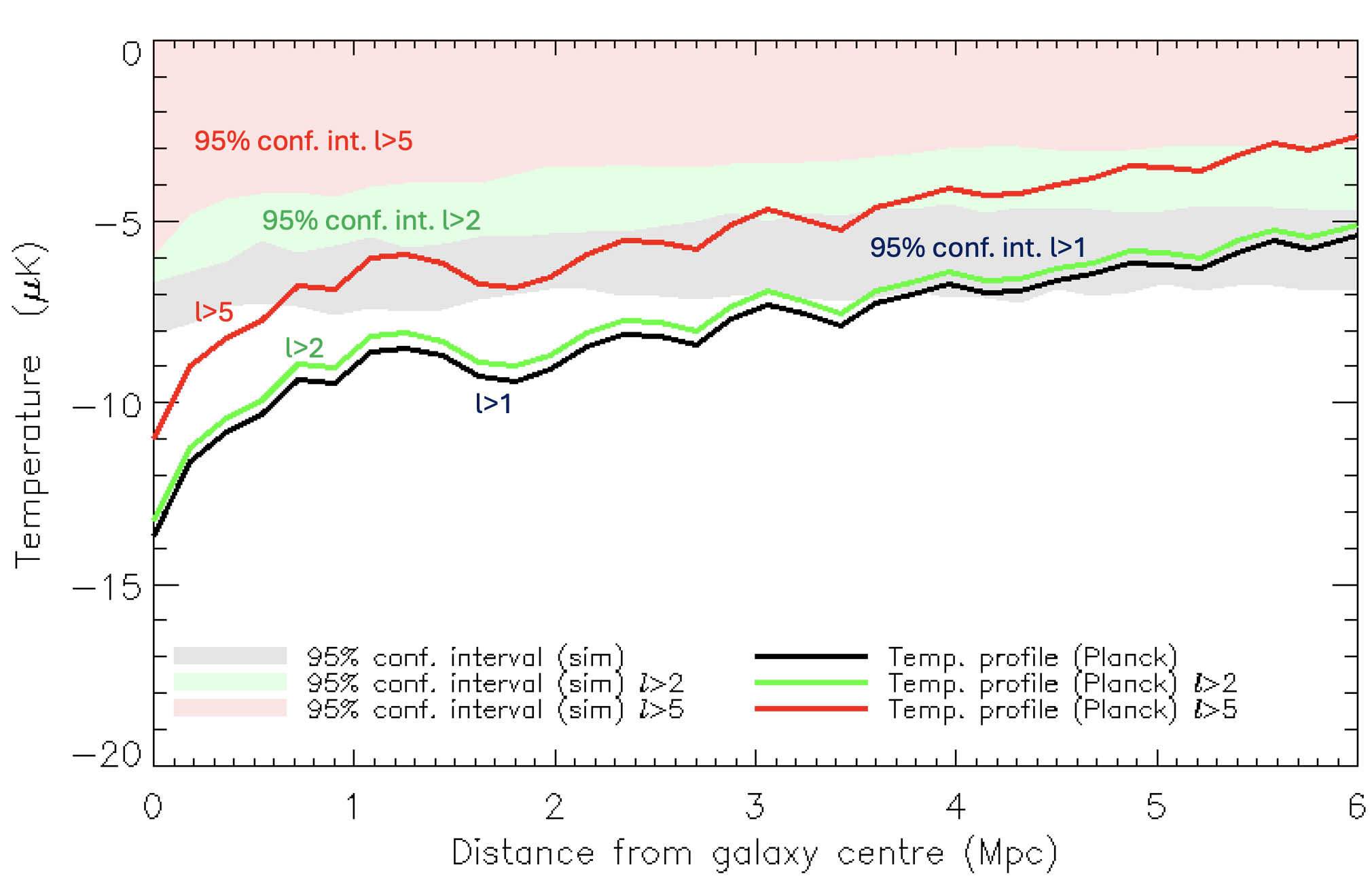}
  \includegraphics[width=\linewidth]{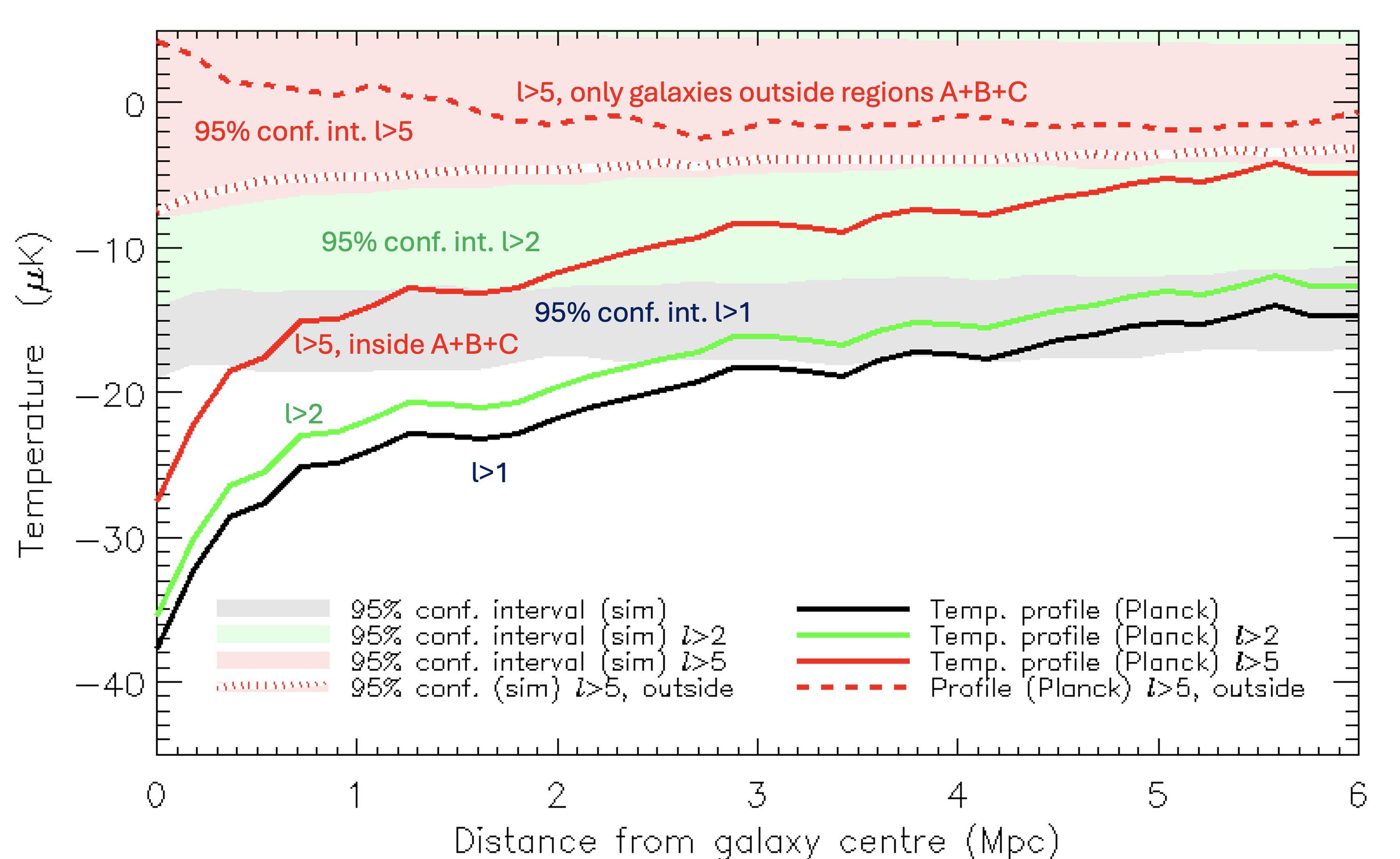}
\caption{The projected radial temperature profile around large ($r>8.5$ kpc) late type spiral galaxies in the redshift range $z=[0.004,0.02]$ (2550 galaxies) when considering the full Planck SEVEM PR4 map (black) line, the map with the quadrupole removed (green line) and the map with the multipoles $\ell\leq5$ removed (red line). The shaded areas show the $95\%$ confidence intervals for the corresponding 600 simulated SEVEM PR4 maps. The lower map shows the same results but limited to the 1250 galaxies in the 3 areas defined in Figure \ref{fig:local}. The profile of the remaining 1300 galaxies outside these areas is shown with a red dashed line with the red-white dashed line indicating the $95\%$ confidence interval. \label{fig:longprof}} 
\end{figure}

\section{Method}
\label{sect:method}

\subsection{Profile depths}
One of our main aims in this work is to study the average CMB temperature around galaxies. In L2023 and A2024, discs with a given angular size around each galaxy was divided in a given number of bands at different angular distances from the galactic centre. The mean temperature within these bands were calculated and averaged over galaxy samples. Due to the large span of redshifts of the galaxies in the samples, causing angular distances of the profiles to represent very different physical distances, we argue that the size of the discs and bands around the galaxies should be calculated in projected distance in Mpc. All profiles in this paper will therefore be shown as a function of Mpc and not degrees.

Previous works have indicated that when looking at galaxies with fewer neighbours (H2023), we find a negative radial CMB temperature profile, with the temperature increasing towards zero Kelvin at a distance of up to $1-2$Mpc from the galactic centre, coinciding with the size of a typical dark matter halo. As most galaxies are not isolated and one will in general see the combined profiles from several neighbouring galaxies, it is difficult to find an exact shape of the radial profile (H2023). In the same work, we found strong indications of a galaxy property dependent temperature profile. In particular, the depth of the profile was found to increase with galaxy size, but the number of isolated galaxies within a given size range was found too small to draw firm conclusions.

Based on the evidence for a radial profile with a large CMB temperature decrement towards the centre of the galaxy, we have chosen to focus our statistical measures solely on the innermost profile bin where we expect to find the strongest signal and which minimizes the overlap with neighbouring galaxies.  This means in practice that we average the CMB temperature over discs of radius $0.2$\;Mpc over all galaxies in the sample, excluding pixels which are masked in the Planck common mask.

In A2024, a $\chi^2$ approach to the significance was attempted where all bins in the temperature profile, also at large distances from the galaxy where no signal is expected, was included. Given that we see a strongly negative profile near the galactic centre, such a statistic will give weaker and weaker significances the more profile bins are included in the $\chi^2$, even with a strong central signal present. The result will therefore depend on the number of bins included in the profile. It was furthermore noted that the bins are strongly correlated. One would need a large number of simulations to obtain a reliable covariance matrix for the $\chi^2$ statistic. The strong correlations indicates that just a very few numbers are sufficient to describe the profile.

We therefore choose one single number which is the {\it profile depth}, defined as the temperature value of the first profile bin covering a disc of radius $0.2$\;Mpc around the galactic centre (although in some simulations where this is a {\it profile height} as the bin temperature can be positive, we will nevertheless refer to the first bin temperature as {\it profile depth}). The value of this profile depth will be averaged over galaxy samples and compared to the temperature value of the corresponding bin in simulations.\\

\begin{figure}[H]
\includegraphics[width=0.975\linewidth]{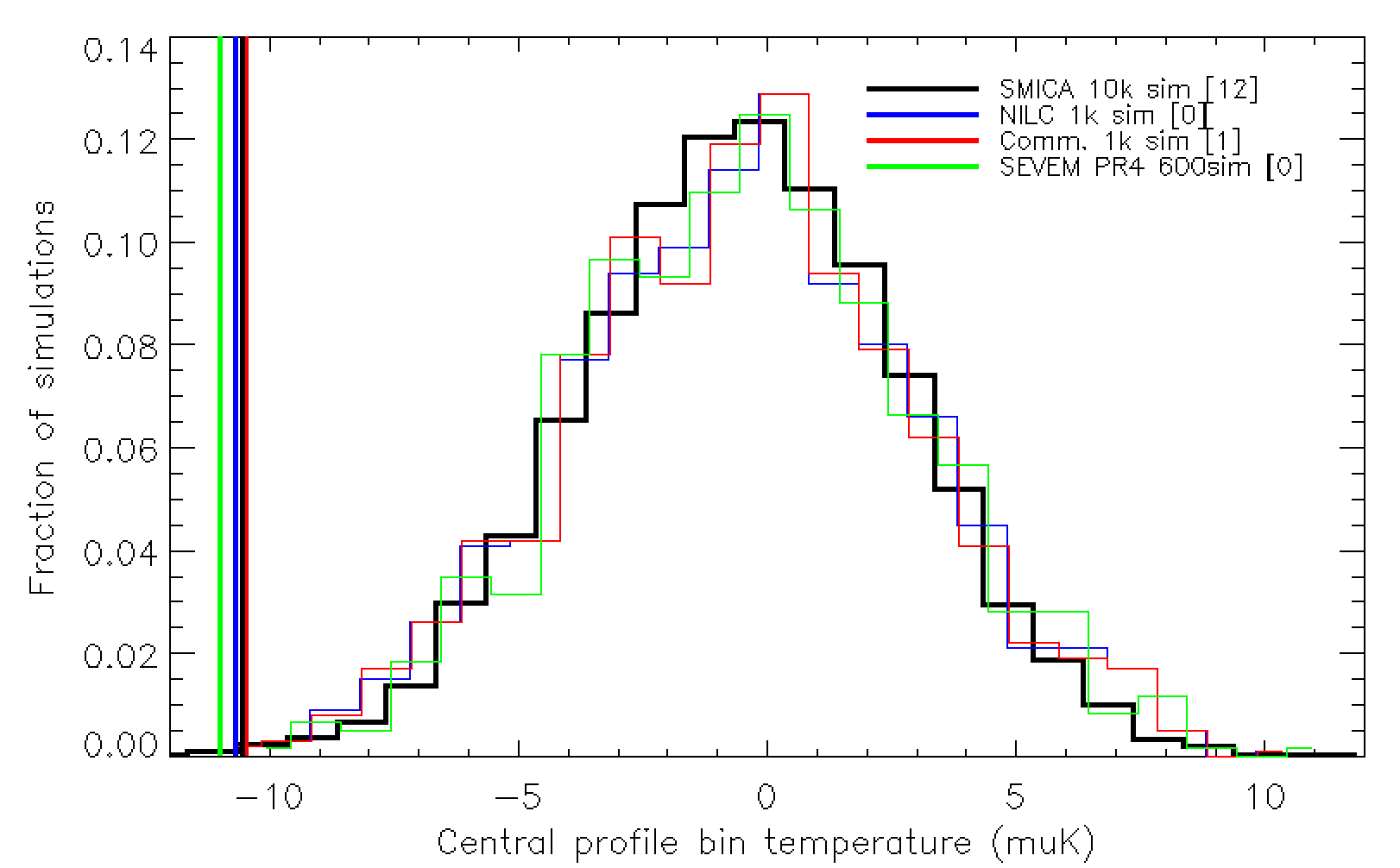}
\includegraphics[width=0.975\linewidth]{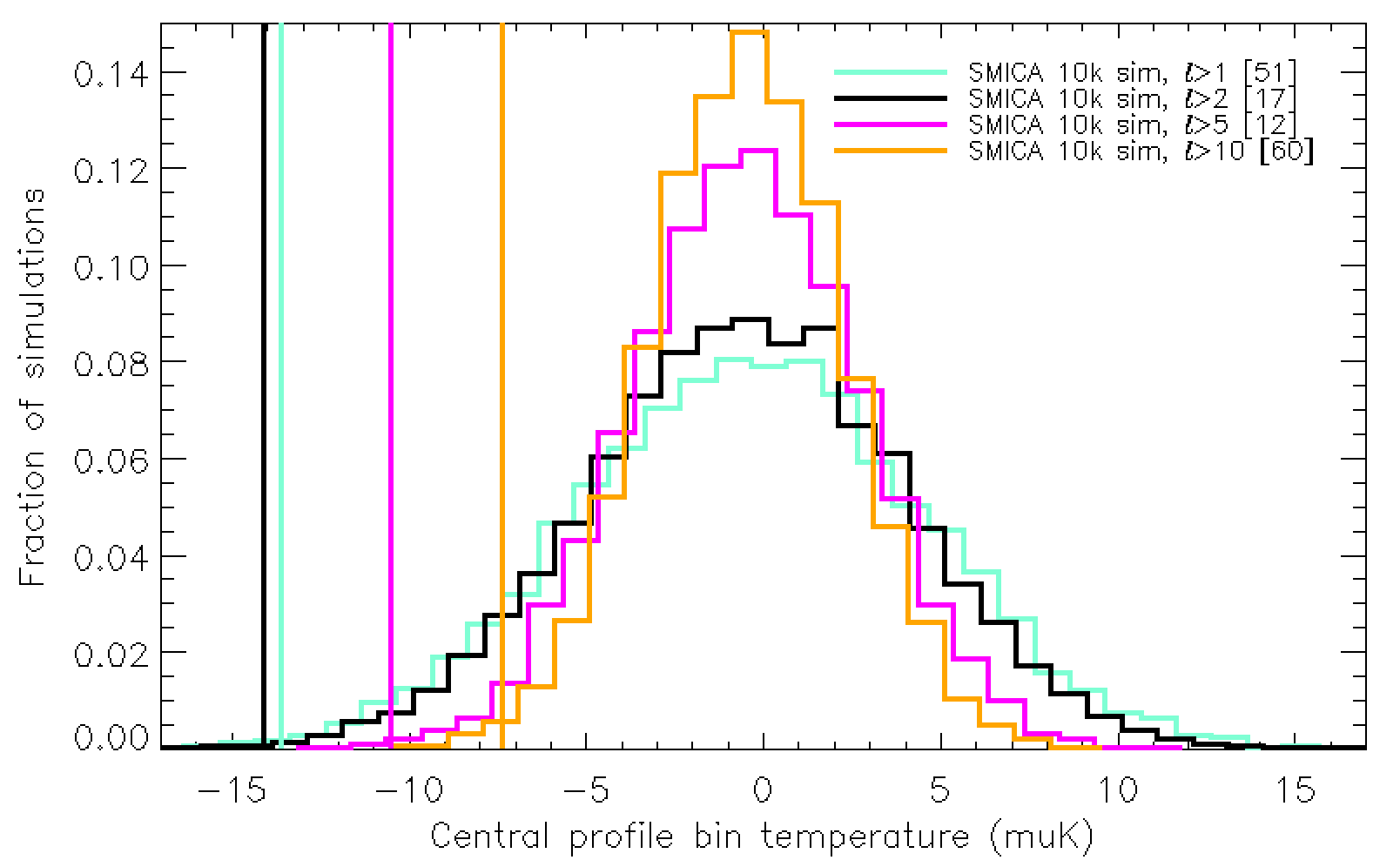}
\includegraphics[width=\linewidth]{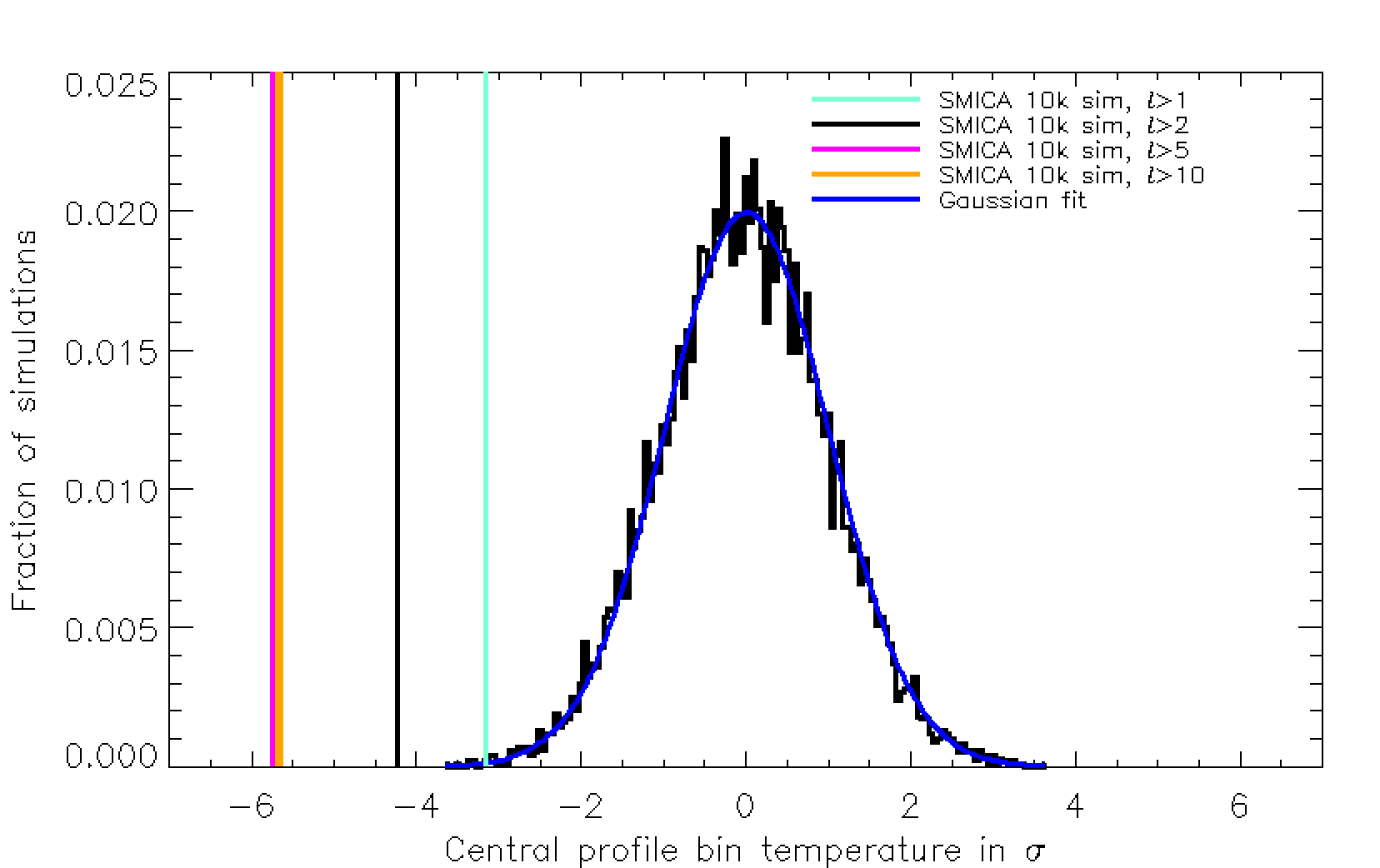}
\caption{The distribution of the first temperature profile bin of Figure \ref{fig:longprof} calculated on simulated maps. In the {\it upper plot}, the $\ell>5$ case for SEVEM PR4 (corresponds to the first bin of the red line in Figure \ref{fig:longprof}) is shown as a green vertical line. The green histogram shows the distribution of this first bin in simulations (corresponding to the first bin of the red shaded area in Figure \ref{fig:longprof}). The corresponding results for the other three Planck foreground subtraction methods (from the Planck PR3 release), SMICA (10.000 simulations), NILC (1000 simulations) and Commander (1000 simulations) are also shown. The {\it middle plot} shows the same results for SMICA (10.000 simulations) only, for the four cases: only mono- and dipole removed, quadrupole removed, $\ell\leq5$ removed and $\ell\leq10$ removed. {\it The lower plot} shows the same as in the middle plot, but for galaxies in dense environments only (defined by the circles in Figure \ref{fig:local}) and with the x-axis in units of standard deviations of the Gaussian fit to the histogram for the $l>2$ case shown as a blue line. In the upper two plots, the number of simulations with a smaller profile depth than the data is shown in brackets. \label{fig:profhist}} 
\end{figure}

Following up the results from H2023, we will investigate the scale dependence of the results. In H2023, we found correlations between the galaxy field and the CMB independently for both the largest angular scales $\ell<5$ as well as the smaller scales. In the following, we will therefore test the presence of the foreground signal independently for the largest and for the smaller scales of the CMB map:
\begin{enumerate}
  \item First we investigate to what extent the results depend on these large scales and test whether the significance of the profile depth persists when removing the best estimate of the low multipole moments. We calculate profile depths of the data map as well as the simulated maps with (1) the best fit quadrupole removed, (2) all multipoles $\ell\leq5$ removed and in some cases (3) all multipoles $\ell\leq10$ removed. In the latter case, the angular scales correspond to about $15^\circ$ to $20^\circ$ on the sky, which is similar to the extension of some galaxy groups/filaments. We therefore expect that the foreground signal around these structures also will be removed when removing multipoles above $\ell\geq10$, possibly reducing parts of the signal. We will test this by gradually removing multipoles until the signal weakens or disappears. 
  \item For the low multipole mode maps which we remove in the previous point, we will, as described below, test the correlation of these largest scales with the galaxy density field.
\end{enumerate}

We will also show that the largest scale fluctuations of the CMB significantly increases the dispersion of profile depths, acting as an additional noise to the profile distribution. If a very large scale temperature fluctuation is superposed on a filament of galaxies, it could easily remove or increase the profile depths for a large number of galaxies. Removing the very first multipoles, the dispersion in the simulations is significantly reduced, increasing the signal-to-noise ratio for the temperature profiles.

\subsection{The look-elsewhere-effect}

Due to the above mentioned galaxy size dependency of the profile depth, we found in L2023 that one has to limit the galaxy sample to galaxies above a certain size limit in order to see a significant foreground signal. The median size corresponding to a radius of $8.5$\;kpc of the 2MRS sample used in that work was the minimum galaxy size used as a basis for the reported detection. Increasing the minimum galaxy size while maintaining a sufficiently large number of galaxies in the sample, the profile depth and significance of the detection increases.

A strong dependence on galaxy morphological type was also observed. Late type spiral galaxies (Sb, Sc and Sd types) were found to give the main contribution to the signal. Our aim here is to compare the profile depth temperature around the large, late spiral galaxies in Planck data with the corresponding temperature calculated on CMB simulations. However, given the number of choices made (such as large vs small galaxies, late spiral galaxies vs early spiral galaxies and elliptical galaxies), it is important to correct for the look-elsewhere-effect: In the data, we looked for the sample of galaxies with the largest detection which we found to be the sample of large late type spiral galaxies. We need to repeat the same procedure for each simulation where we look at the same possible samples (such as large/small and different galaxy type) and record the profile from the sample with the largest detection to compare with the largest detection in the data.

We will apply the following procedure for each CMB simulation and also for the Planck data to correct for the look-elsewhere-effect:
\begin{enumerate}
  \item We calculate the average profile depth over galaxies in different galaxy samples defined by possible combinations of a set of galaxy parameters such as small/large (compared to the median size $8.5$kpc) galaxies, elliptical/early/late spiral galaxies, different maximum redshifts or different minimum sizes. We obtain one profile depth temperature $T_{si}$ for each sample $s$ and simulation $i$. We require each sample to have a minimum of 1000 galaxies in order to avoid small samples with large fluctuations, but we reduce this minimum requirement when considering smaller parts of the sky with less galaxies.
  \item We then calculate the standard deviation $\sigma_s$ for each galaxy sample $s$ of the profile depths $T_{si}$ over simulated CMB maps $i$.
  \item In order to be able to compare the profile depths $T_{si}$ between different galaxy samples $s$ with a different number of galaxies in each sample, for each simulation/data $i$ and sample $s$, we convert the profile depth temperature $T_{si}$ to the number of standard deviations (calculated in the previous point) away from zero temperature as $T_{si}/\sigma_s$
\item For each simulation/data, we record the value of the profile depth (in standard deviations) $|T_{si}|/\sigma_s$ for the sample $s$ with the largest deviation (in absolute value) for the given simulation.
  \item We plot the histogram of the recorded maximum (absolute value) profile depths (measured in standard deviations) for all the simulations and compare to the data. Our measure of significance is then the number of simulations with a larger deviation than the data.
\end{enumerate}

\begin{figure}[htbp]
  \includegraphics[width=\linewidth,page=1]{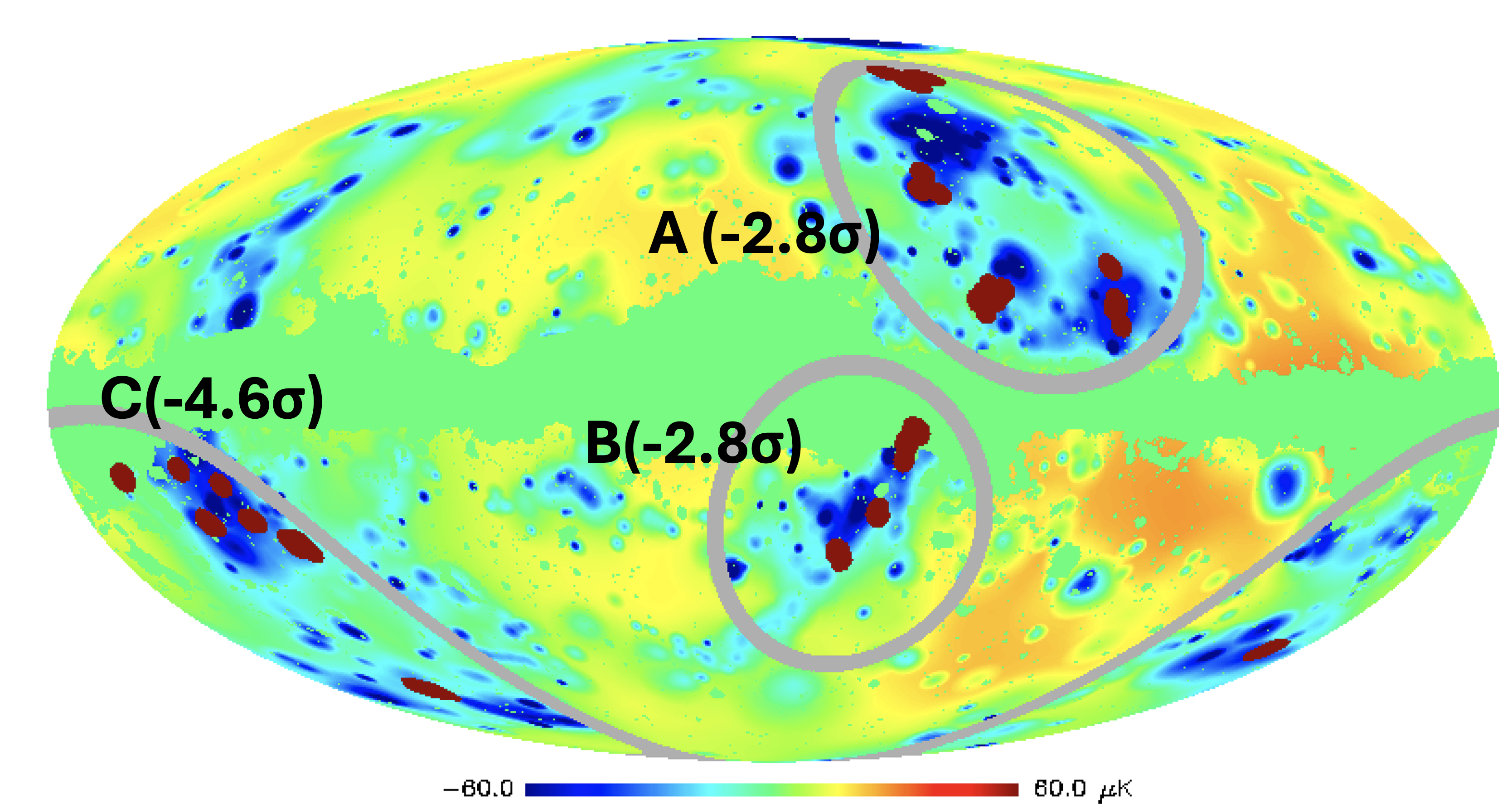}
  \includegraphics[width=\linewidth,page=1]{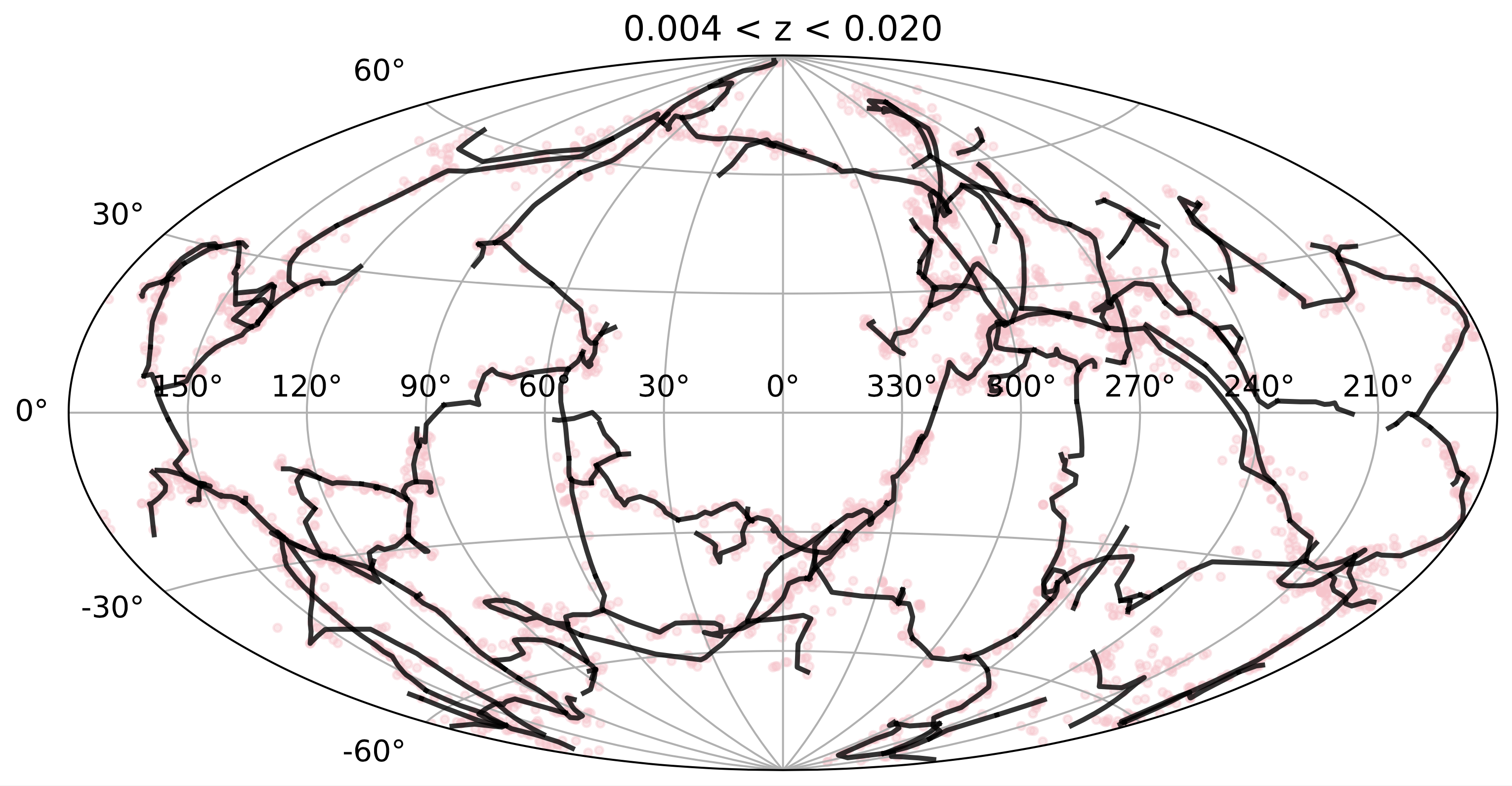}
  \includegraphics[width=\linewidth,page=1]{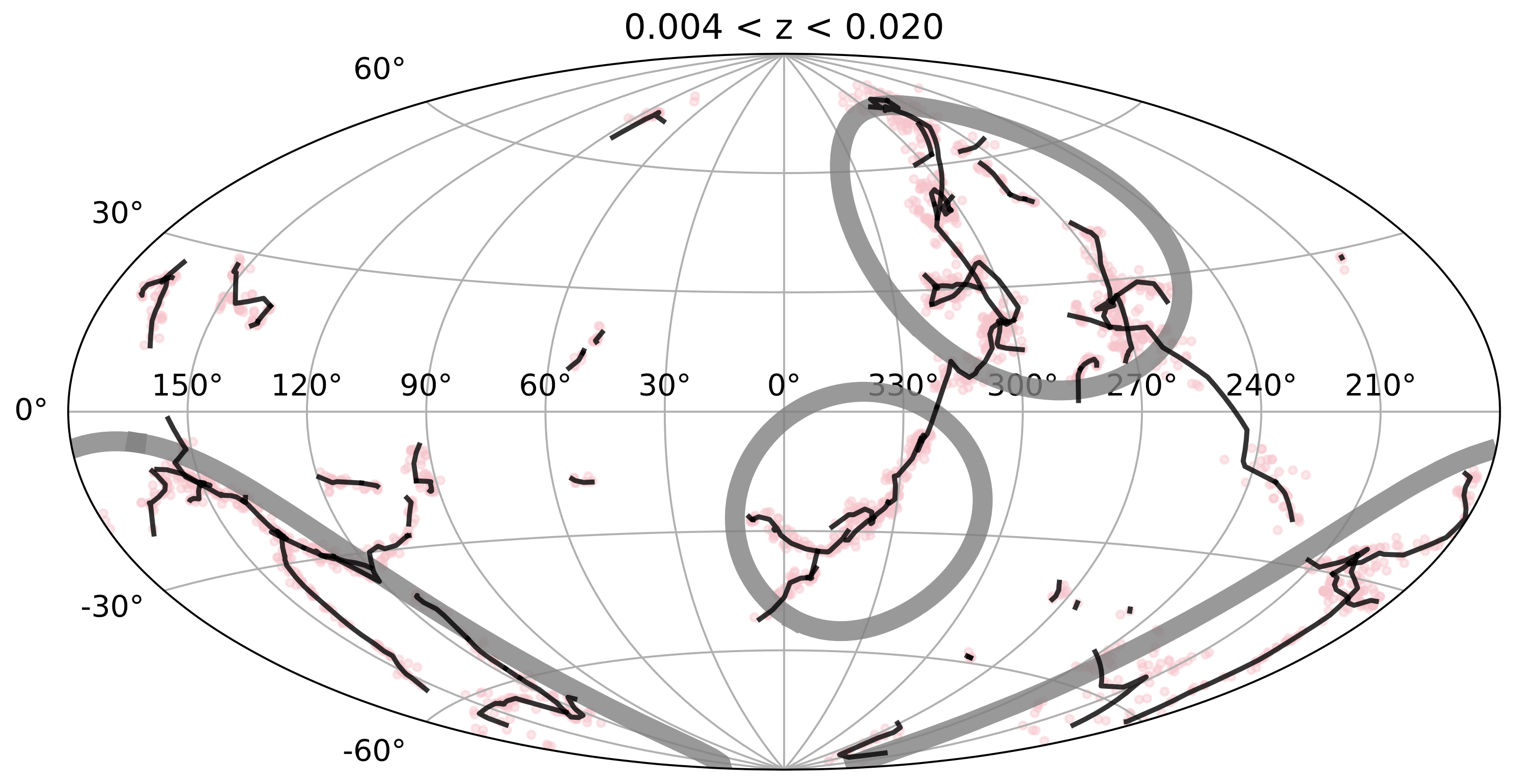}
  \caption{Three circular areas around the extended filaments with the highest galaxy density, denoted area A, B and C.  {\it Upper plot:} the map shows the galaxy foreground temperature model from H2023. The model shows high galaxy density of nearby ($z<0.017$) galaxies as low temperature. Red dots show the positions of massive ($>10^{13}M_\odot$) galaxy groups containing large spiral galaxies within $z<0.02$.  The significance of the profile depth in Planck SMICA maps with $\ell\leq5$ removed for spiral galaxies within the given zone is quoted in standard deviations calibrated on 10.000 simulations. {\it Middle and lower plot:} The projected distribution of filaments for the redshift range $z \in [0.004, 0.02]$ for both the total sample (middle panel) and the highest linear K-band luminosity density (lower panel). The dotted pink lines represent 2MRS galaxies with a perpendicular distance of less than $4\,\, \mathrm{Mpc}$ to the filaments. \label{fig:local}}
\end{figure}

\subsection{Low multipole correlations}

Finally, we follow up on the results of H2023 where we reported a significant correlation between the largest scales ($\ell\leq5$) of the Planck CMB map and the galaxy distribution map where each galaxy is represented with a simple linear foreground model as detailed in H2023. In order to reduce correlations with the smaller scales in the estimation of the individual large scale multipole maps, we downgrade the resolution of the CMB map as well as the galaxy model map to $N_\mathrm{side}=4$. We apply the galactic common mask before degrading and extend the mask correspondingly. We use a $\chi^2$ minimization approach to estimate all the individual $\hat a_{\ell m}$ from $\ell=0$ to $\ell=4 \cdot N_\mathrm{side}=16$:
\[
\chi^2 = \sum {\bf d}^\dagger C^{-1}{\bf d}
\]
where the elements of the data vector is given as
\[
d_{\ell m} = \sum_{\Omega}\hat a_{\ell m} Y_{\ell m}(\Omega) M(\Omega)- \tilde a_{\ell m}
\]
where the sum is performed over all pixels of the map and $\tilde a_{\ell m}$ are the pseudo-$a_{\ell m}$ with the extended mask $M(\Omega)$. The coupling matrix is calculated as
\[
C_{\ell m,\ell' m'}=\sum_{\Omega}M(\Omega)Y_{\ell m}(\Omega)Y_{\ell' m'}(\Omega)
\]

From the estimated $\hat a_{\ell m}$, we create multipole maps for each $\ell$ as
\[
\hat M_\ell(\Omega) = \sum_{m=-\ell}^\ell \hat a_{\ell m}Y_{\ell m}(\Omega)
\]
In this way, we obtain $\hat M_\ell(\Omega)$ for both the CMB map and the galaxy model map up to $\ell=4 \cdot N_\mathrm{side}=16$. We use a simple model map where we construct a linear foreground profile around each large, late type spiral galaxy with a profile depth of $-10\mu$K and a radius of $1$Mpc. We know that this simple model is not accurate. It does for instance not take into account possible galaxy size dependency and galaxy environment density as in H2023. For this reason, we are only interested in looking for correlations between the shape of the galaxy distribution and the CMB map, not the amplitude. We therefore normalize each $\hat M_\ell(\Omega)$ by its standard deviation, obtaining $M_\ell^\mathrm{norm}(\Omega)$ and construct a correlation coefficient up to a given $\ell_\mathrm{max}$ as
\[
c_{\ell_\mathrm{max}}=\sum_{\ell=2}^{\ell_\mathrm{max}}\sum_{\ell'=2}^{\ell_\mathrm{max}} \sum_\Omega M_{\ell,\mathrm{CMB}}^\mathrm{norm}(\Omega)M_{\ell',\mathrm{galaxies}}^\mathrm{norm}(\Omega)
\]
taking into account correlations and cross-correlations between the maps. We will compare the histogram of these correlation coefficients for the simulations to the corresponding correlation coefficients for the data, using different values of $\ell_\mathrm{max}$.

The low-$\ell$ multipole maps estimated from this procedure are also the same ones used to subtract from CMB maps in order to study scale dependence and increase signal-to-noise for radial profiles as described above. For removing multipoles above $\ell>16$, this procedure becomes numerically unstable due to the lack of small scale information outside the mask to estimate $a_{\ell m}$ for higher multipoles. In this case, we apply a much simpler procedure which only can remove larger multipole ranges without estimating individual $a_{\ell m}$. We smooth and downgrade the map to low resolution, zero $a_{\ell m}$ for higher multipoles and subtracting the resulting large scale map outside the mask. For the purpose of removing large scales, this simplified procedure is found to give almost identical results to the more exact procedure in overlapping multipole ranges. We will quantify the error in the profile depth due to the uncertainty in the low-$\ell$ removal procedure by comparing to the profile depth in simulations where exact low-$\ell$ removal is possible. In Figure \ref{fig:lremoved}, we see the Planck SMICA map with the largest scales removed which will be used for calculating profile depths.

\begin{figure}[htbp]
\includegraphics[width=\linewidth]{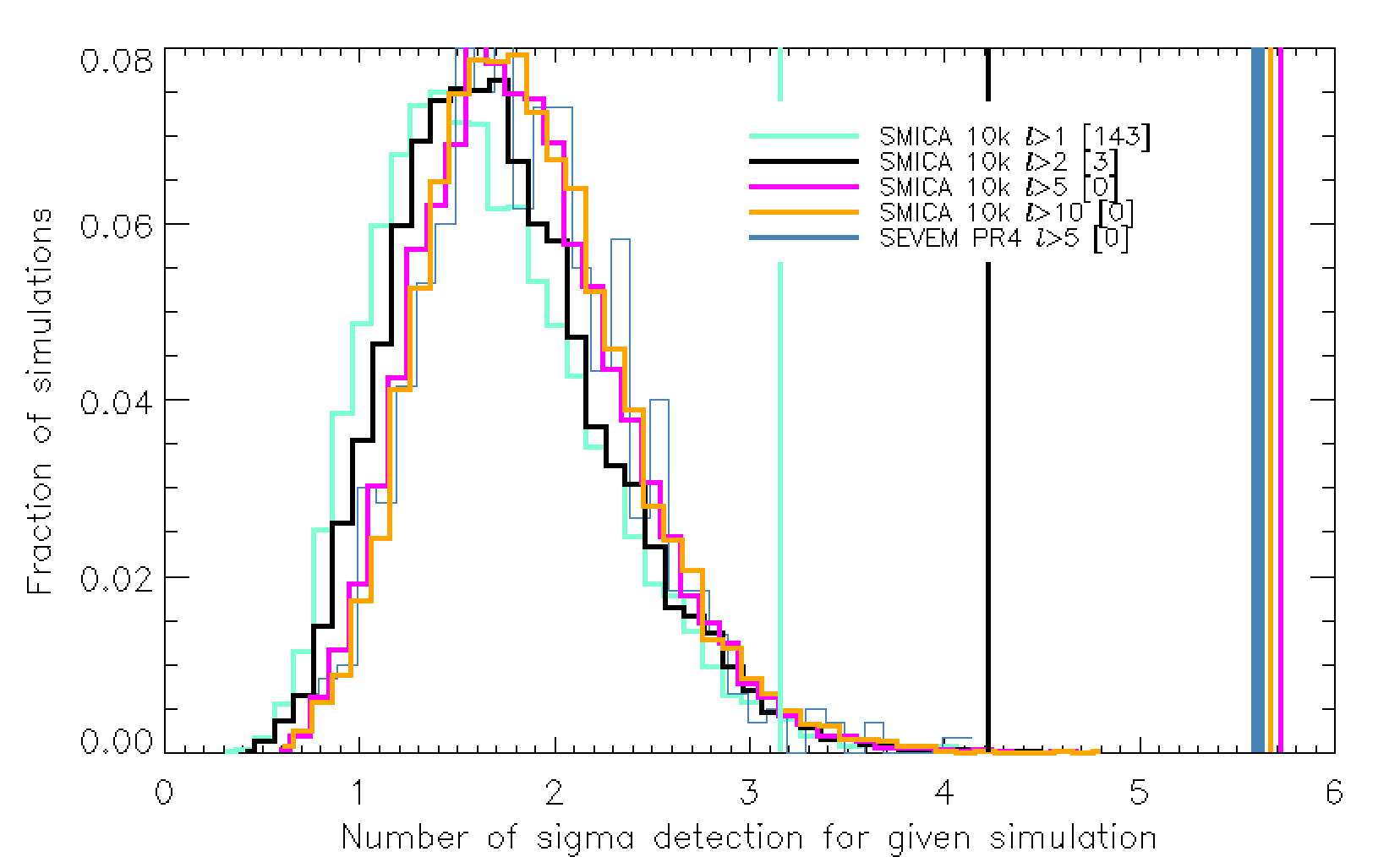}
\caption{As in Figure \ref{fig:profhist}, we show the distribution of the first temperature profile bin, but now measured in the number of $\sigma$ deviations from the mean temperature value. For each simulated CMB map, the temperature of the first profile bin is calculated for the combinations of the samples of smallest/largest galaxies, early/late spirals/elliptical galaxies and galaxies outside/inside dense galaxy filaments. Of these combinations, the value of the combination which maximizes the absolute value deviation from the mean profile for each given simulation is plotted in the histogram. The vertical lines show the deviations obtained in the same manner for the Planck data. The case SMICA with 10.000 simulations for the four cases: only mono- and dipole removed, quadrupole removed, $\ell\leq5$ removed and $\ell\leq10$ removed are shown as well as SEVEM PR4 with  $\ell\leq5$ removed. The number in brackets show the number of simulations with a higher deviation than the data. \label{fig:devhist0}} 
\end{figure}

\begin{figure}[htbp]
\includegraphics[width=0.985\linewidth]{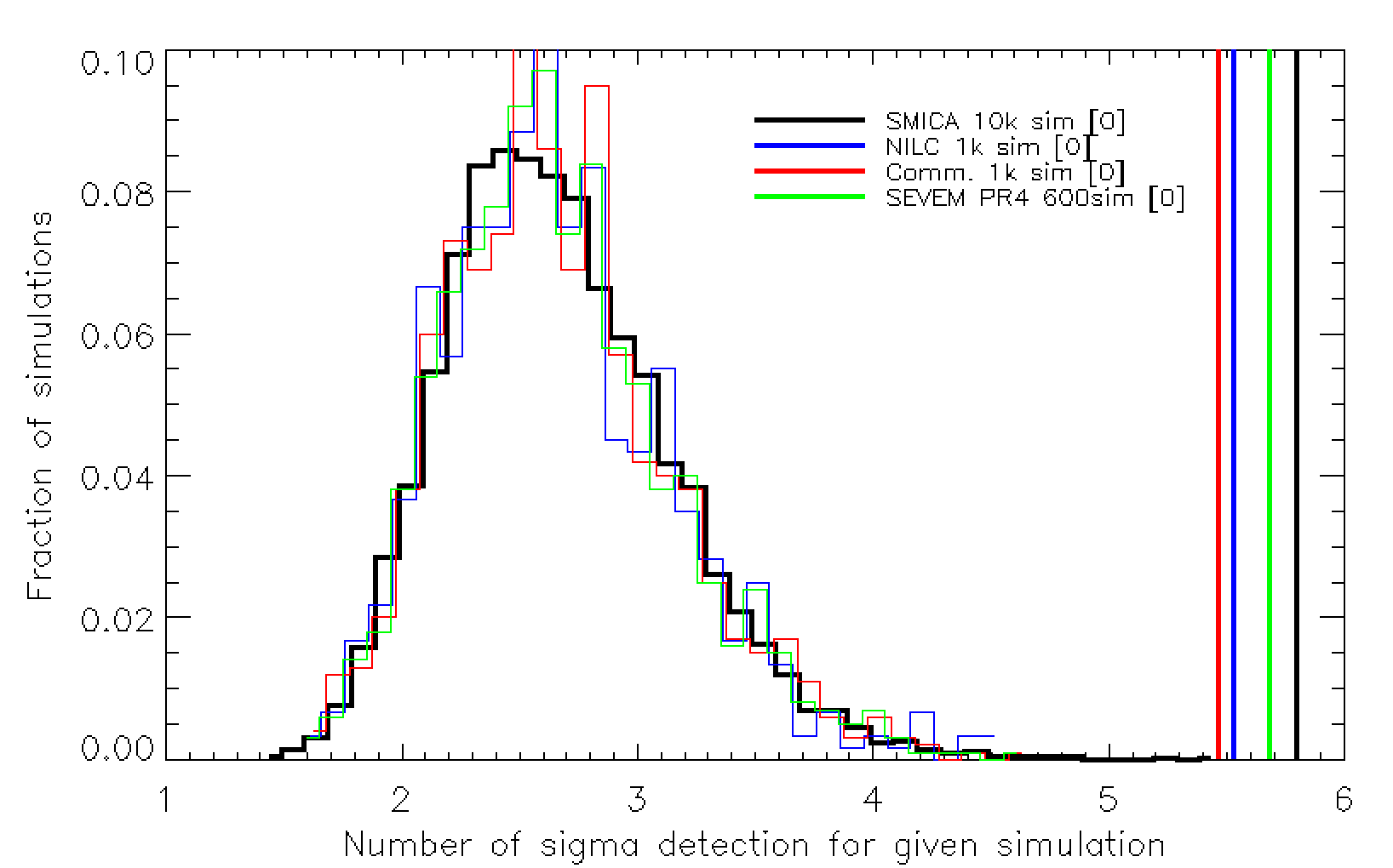}
\includegraphics[width=\linewidth]{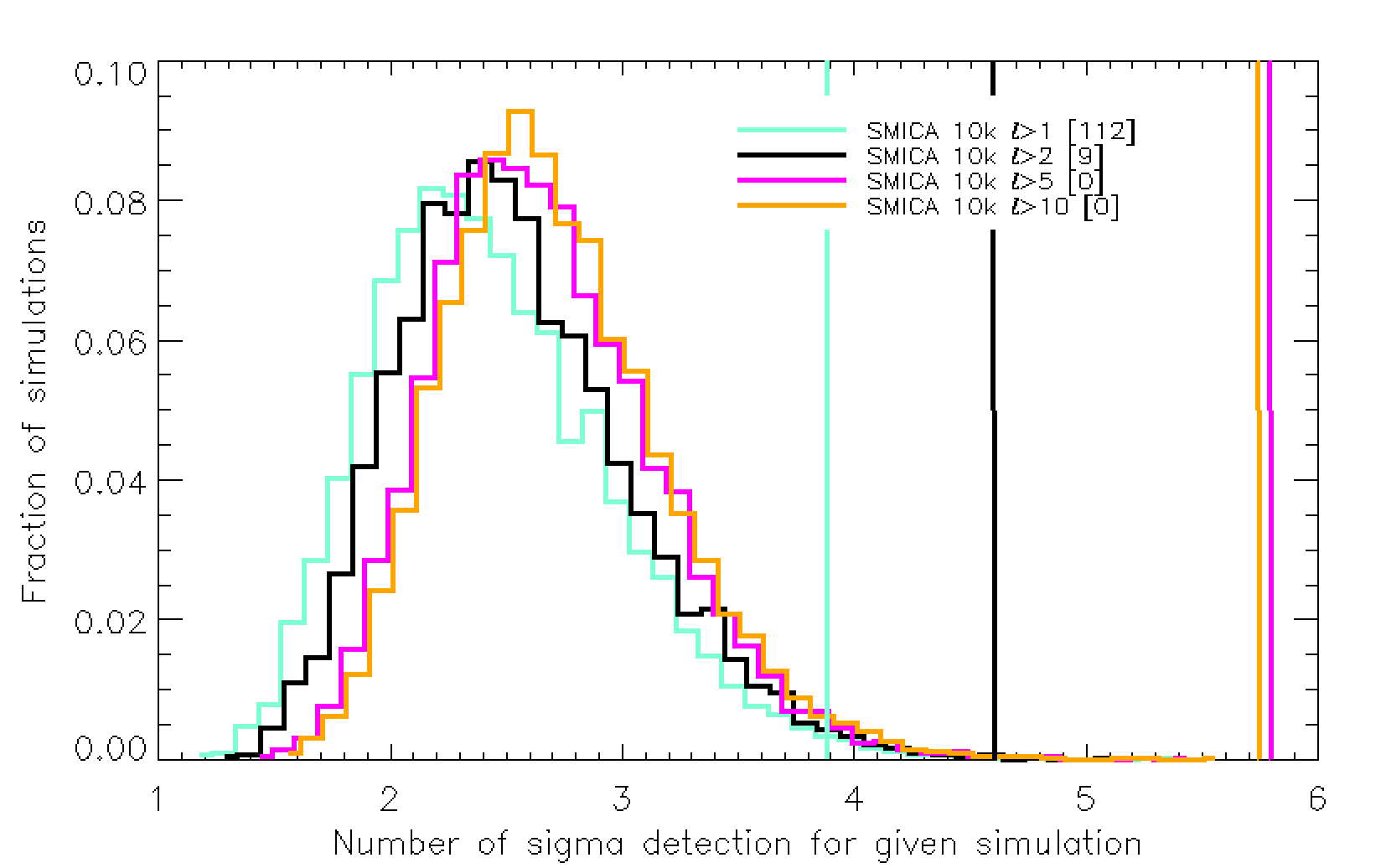}
\caption{As in Figure \ref{fig:profhist}, we show the distribution of the first temperature profile bin for galaxies in massive filaments, but now measured in the number of $\sigma$ deviations from the mean temperature value. For each simulated CMB map, the temperature of the first profile bin is calculated for all galaxy samples with the combinations of maximum redshifts up to $z<0.25$ in steps of $\Delta z=0.01$ and all possible minimum sizes in steps of $0.5$ kpc, early/late spirals/elliptical galaxies and galaxies outside/inside dense galaxy filaments. The sample of galaxies with the combination of maximum redshift, minimum galaxy size, galaxy type and inside/outside filaments which maximizes the deviation (positive or negative deviation) from the mean profile depth for each given simulation is used and the corresponding deviation plotted in the histogram. The vertical lines show the deviations obtained in the same manner for the Planck data. The {\it upper plot} shows results for maps with multipoles $\ell\leq5$ removed for different foreground subtraction methods and the lower plot shows only SMICA results for the four cases: only mono- and dipole removed, quadrupole removed, $\ell\leq5$ removed and $\ell\leq10$ removed.  The number in brackets show the number of simulations with a higher deviation than the data. \label{fig:devhist}} 
\end{figure}

\section{Results}
\label{sect:results}

\subsection{Profile depths}
We firstly explore the sample of galaxies used in L2023, H2023 and GL2024, where late spiral galaxies with radius larger than $8.5$\;kpc (the median size of the 2MRS sample used in L2023, note that $H_0=100$km/s/Mpc was used there to convert angular size to physical size, here we adopt the same convention), were found to give the largest foreground signal in the CMB. Whereas these papers focused on $z<0.015$ and $z<0.017$, we will here extend our sample to $z<0.02$ giving a total of about 2500 galaxies. Beyond this redshift, the uncertainty of galaxy type identification increases substantially as can be seen in Figure 14 of \cite{2mrs} where there is an unexpected sharp decline in the observed relative fraction of spirals. Given the assumption, which we discuss in more detail below, that spiral galaxies dominate the signal, good morphological identification is important for correctly identifying the foreground signal. Note that a different interpretation of the decline in spiral galaxies is that the spiral fraction locally is considerably higher than the average in the Universe. In this case, we would still expect a higher foreground signal for nearby structures and the $z<0.02$ is still a good limiting redshift for this study.

As in H2023 and GL2024, we will also limit our sample to redshift larger than $z>0.004$ as peculiar velocities may strongly affect the distance to the most nearby galaxies. Also the large angular areas that the foreground signal from these galaxies occupy would contribute significantly to the uncertainties in the profiles.\\

In the upper plot of Figure \ref{fig:longprof} we see the mean radial temperature profile around these galaxies out to $6$Mpc from the galactic centre for 3 different low-$\ell$ multipole cuts using SEVEM PR4 data. The corresponding 95\% confidence intervals based on the 600 SEVEM PR4 simulations are shown for each cut. The plot confirms profiles shown in other works (L2023, H2023, GL2024 and C2024) and also shows the reduced spread in the simulated profiles when the lowest multipoles are removed.  We can see that for the maps with the quadrupole removed as well as the case with $\ell \leq 5$ removed, the profiles are well outside the 95\% confidence interval out to several Mpc and actually follow closely the 99.7\% confidence interval (not shown), but the individual bins of the profiles are highly correlated. Note that the difference between the profiles for the full map and for the map with the quadrupole removed is very small because of the small value of the quadrupole in the Planck CMB map. However, the distribution of these profiles in simulations shows that the maps with the quadrupole removed have much less scatter. For this reason, the detection of the foreground is much more significant for the case with the quadrupole removed.\\

As explained above, here we will focus on the first bin of the profile, covering a radius of $0.2$Mpc around the galaxy spanning an angular radius of about $0.25^\circ$ at $z=0.01$. The two upper plots of Figure \ref{fig:profhist} shows the distribution of the temperature of this first bin in simulations as well as in the data. Results from maps based on different $\ell$-cuts are shown in the middle plot and the case with $\ell\leq5$ removed for all 4 Planck foreground subtraction methods are shown with their respective simulations in the upper plot. We see that in all cases the detection is about $3\sigma$ with, depending on the $\ell$-cut, only $0.0012\%$ to $0.006\%$ of the simulations showing a larger detection. In the middle plot, we see clearly that the width of the distribution of profile depths is getting narrower when more large scale multipoles are removed, increasing the signal-to-noise for the profile depth which depends on smaller scale fluctuations. Clearly the foreground signal not only persists when removing the largest scales, but its significance even increases showing that the a significance part of the signal originates from smaller scale fluctuations $\ell>10$.\\

In L2023 it was shown that galaxies in dense environments generally show a stronger signal. We will now focus on the case with the strongest detection when $\ell\leq5$ is removed, in order to investigate if the signal is localized to a certain part of the sky or if all galaxies in different parts of the sky show a similar signal. Figure \ref{fig:local} shows the foreground model map of H2023. Cold areas show the largest density of late spiral galaxies. We see that there are three areas in the sky where the nearby galaxies are densely clustered in filaments.  We name these areas A, B and C. The red dots indicate the positions of all the most massive galaxy groups within $z<0.02$ and we see that their positions coincide with these areas. When looking at the projected filaments of nearby galaxies, focusing on filaments with linear K-band luminosity density above the mean value ($\sim 10^{11.75}\,\, \mathrm{Mpc^{-1}}\,\, \mathrm{L_{\odot}} $), we see that these denser structures are mainly concentrated in these three circular regions in contrast with the low density filaments which cover homogeneously the whole sky.

We have calculated the profile depth of late spiral galaxies in each of these areas. The upper plot of Figure \ref{fig:local} shows the significance of the profile depth in each area for the case with $\ell\leq5$ removed. Table \ref{tab:local} presents the local significance for different cases as well as the total significance considering all the 3 areas of densely clustered spiral galaxies. The profile depth averaged only over all large spiral galaxies outside these areas shows no significant foreground signal. As a measure of galaxy density, we find that the mean distance to the 5th neighbour galaxy is on average $1.7\pm1.1$\;Mpc within areas A+B+C whereas for galaxies outside these areas, the mean distance to 5th galaxy of $2.5\pm1.6$\;Mpc.

We see that when considering only the galaxies within dense filaments (a total of about 1250 galaxies, half of the total number of large spirals in this redshift range), the significance of the profile depth increases to $5.7\sigma$ when the largest scales have been removed. The histogram of the corresponding profile depths in the 10.000 SMICA simulations is shown in the lower plot of Figure \ref{fig:profhist} with a Gaussian fit showing that the distribution is close to Gaussian, validating the use of standard deviations as a measure of significance. We can see from Table \ref{tab:local} that the three areas show significantly negative profiles, but area C is by far the most significant area. We find that galaxies in the Cold Spot area are contributing substantially to the high significance of area C. In GL2024, it was found that the Cold Spot shape to a large degree can be reproduced considering a foreground signal from the galaxies in this area. Not considering galaxies within a radius of 10 degrees around the Cold Spot, the significance for the case with $\ell\leq5$ removed, drops from $4.8\sigma$ to $2.3\sigma$ in area C, similar to the area A and B, and the total significance for the whole area A+B+C falls from $5.7\sigma$ to $4.4\sigma$. The contribution from Cold Spot galaxies is important, but the signal is still highly significant when these are excluded and the signal is then evenly distributed around the filaments of high galaxy density. The full profiles, considering only the galaxies either within or outside the areas A+B+C, are shown in the lower plot of Figure \ref{fig:longprof}.

The simulated maps used to calibrate the standard deviation are based on the Planck best fit $\Lambda$CDM model. The power spectrum for small multipoles $\ell<30$ and in particular the quadrupole, are well known to be lower than expected in this model (See \cite{planck_powspec_2013} and references therein). The difference between the actual power spectrum and the simulated ones based on the standard $\Lambda$CDM model could therefor give rise to a biased distribution of profile depths. We therefore produced a set of simulations based on the best fit Planck power spectrum (and not the best fit model), combined with the Planck noise simulations. We tested two of the most significant cases and found that for $\ell>1$ using galaxies inside the areas A+B+C, the significance {\it increased} from $3.18\sigma$ to $3.7\sigma$. Similarly for the $\ell>5$ case, the significance increased from $5.7\sigma$ to $5.9\sigma$. Taking the above results and consideration into account, this is easily understood: the Planck power spectrum has {\it lower} low-$\ell$ power than the simulations based on the model. Smaller large scale structure reduces the scatter of small-scale profile depths in the same manner as when removing the multipoles. We conclude that the difference between simulated maps and the real data due to the low-$\ell$ anomaly gives rise to a slightly underestimated significance.

In order to further study the scale dependence of the effect, we will focus on the most significant case, and remove multipoles in intervals from $\ell<16$ to $\ell<1024$ using the simplified procedure described above. In table \ref{tab:remove_many} we show the results. We can see that the foreground is strongest at lower multipoles $5<\ell<128$, but present at the $2.5\sigma$ level even at $\ell>1024$, even though the uncertainty due to multipole removal is larger. At this scale we are approaching the scale of the inner bin of $0.2$\,Mpc for which we measure the profile depth. The rapidly increasing error of the multipole removal at smaller scales is due to uncertainty around the point source mask. We conclude that the signal seems present, both at the scale of the cosmic filaments around the galaxies, but also to some extent at the scales of the individual galactic halos.

\begin{table*}[htbp]
  \begin{center}
    \caption{Significances in areas A, B and C for the SMICA map.}
    \label{tab:local}
    \begin{tabular}{c|c|c|lr|lr|lr|lr}
      \textbf{galaxies in...} & \textbf{centre $\ell, b$} & \textbf{radius} & \textbf{Full map}& & \textbf{$l>2$}& & \textbf{$l>5$}&  & \textbf{$l>10$}&\\
\hline
Area ABC & NA & NA  & $8/10^4$ & {\bf($\bm{-3.18\sigma}$)} & $0/10^4$ & {\bf ($\bm{-4.23\sigma}$)}  &  $0/10^4$ & {\bf ($\bm{-5.72\sigma}$)}  &  $0/10^4$ & {\bf($\bm{-5.67\sigma}$)}\\
Area A & $290^\circ$, $40^\circ$ & $35^\circ$  & $639/10^4$ &{\bf($\bm{-1.53\sigma}$)}  &  $323/10^4$ &{\bf($\bm{-1.87\sigma}$)}  &  $27/10^4$ &{\bf($\bm{-2.79\sigma}$)}  &  $19/10^4$ &{\bf($\bm{-2.99\sigma}$)}\\
Area B & $340^\circ$, $-25^\circ$ & $30^\circ$  & $118/10^4$ &{\bf($\bm{-2.29\sigma}$)}  &  $39/10^4$ &{\bf($\bm{-2.73\sigma}$)}  &  $30/10^4$ &{\bf($\bm{-2.77\sigma}$)}  &  $20/10^4$ &{\bf($\bm{-2.88\sigma}$)}\\
Area C & $170^\circ$, $-45^\circ$ & $40^\circ$  & $142/10^4$ &{\bf($\bm{-2.14\sigma}$)}  &  $1/10^4$ &{\bf($\bm{-3.63\sigma}$)}  &  $0/10^4$ &{\bf($\bm{-4.58\sigma}$)}  &  $1/10^4$ &{\bf($\bm{-4.28\sigma}$)}\\
\hline
    \end{tabular}
  \end{center}
  \tablefoot{We show the significances of the profile depth for galaxies in the areas A, B and C (see Figure \ref{fig:local}) as well as the total significance over all three areas. Significance is quoted as the number of SMICA simulations (among 10.000) with a lower profile depth and the number of standard deviations in parenthesis, assuming a Gaussian distribution.}
\end{table*}

\begin{table*}[htbp]
  \begin{center}
    \caption{Profile depth in areas A+B+C for larger multipole cuts.}
    \label{tab:remove_many}
    \begin{tabular}{|c|c|c|c|}
      \textbf{$\ell$-cut} & \textbf{distance from zero temp.} & \textbf{$\ell$-cut} & \textbf{distance from zero temp.}\\
      \hline
      $\ell>5$ \textrm{(full procedure)} & $-5.7\sigma\pm0.4\sigma$&&\\
      \hline
      $\ell>5$   & $-5.9\sigma\pm0.4\sigma$  &  $\ell>128$   & $-2.5\sigma\pm0.2\sigma$\\
      $\ell>16$  & $-5.1\sigma\pm0.2\sigma$  &  $\ell>256$   & $-3.1\sigma\pm0.3\sigma$\\
      $\ell>32$  & $-4.5\sigma\pm0.2\sigma$  &  $\ell>512$   & $-1.2\sigma\pm0.6\sigma$\\
      $\ell>64$  & $-3.5\sigma\pm0.2\sigma$  &  $\ell>1024$  & $-2.5\sigma\pm1\sigma$\\
      \hline
    \end{tabular}
  \end{center}
  \tablefoot{Profile depth in areas A+B+C for large galaxies in the SMICA map. The profile depth is measured in terms of standard deviations away from zero temperature for the given multipole cut. If not otherwise stated, the multipoles have been removed using the simplified method described in the text. The errors in the profile depth is due to the uncertainty in the multipole removal.}
\end{table*}

\subsection{Correcting for the look-elsewhere-effect}

We now correct for the look-elsewhere-effect by, for each simulation, recording the largest (positive/negative) mean profile depth found when considering possible combinations of samples with (1) small/large galaxies, (2) type of galaxy (early/late/elliptical) and (3) galaxies close or distant to dense filaments (the 3 zones in Figure \ref{fig:local}). As described above, it varies from simulation to simulation which of these samples gives the largest (absolute) profile depth and we optimize the sample configuration for each simulation and compare the largest (absolute) profile depth with that found in Planck data. For Planck data, the optimal configuration is, as described above, late type large spiral galaxies inside the 3 regions of high galaxy density.

The results are shown as a histogram of deviations in Figure \ref{fig:devhist0}. The recorded profile depth for each simulation is measured in the number of standard deviations for the given sample considered.  A smaller sample of galaxies, such as one with elliptical galaxies will have a larger spread of mean profile depths due to the increased statistical uncertainty. Measuring profile depth in units of standard deviations from zero Kelvin calibrated on simulations for the given sample instead of in units of temperature, we obtain numbers which can be compared between the different samples. We see in the figure that even when correcting for all these factors, the significance remains above the $3-4\sigma$ level for the maps with the largest scales removed as none of the 10.000 simulated maps show similarly large absolute profile depth for the individual optimal samples.\\

We repeat the previous exercise, but now extending the space of possible parameters by also allowing the size and redshift ranges to vary. For each simulation, we also found the combination of maximum redshift and minimum galaxy size which gave the largest temperature deviation in the first bin for the given simulation. Note that we restrict the maximum possible redshift to $z<0.25$ beyond which the galaxy classification is highly uncertain.  The maximum deviation for the simulations as well as the data is shown in the histogram of Figure \ref{fig:devhist}. Again, we first show the case where the multipoles $\ell\leq5$ have been removed for all 4 foreground subtraction methods. Then we show for SMICA, 4 different multipole cuts. We find again up to $3-4\sigma$ detection depending on the number of multipoles removed, even when correcting for optimal redshift and galaxy size in each simulation. For the data, we found for all foreground subtraction methods that the optimal minimum galaxy size is $r>8.5$ kpc and maximum redshift $z<0.019$.\\

\begin{figure}[htbp]
  \includegraphics[width=\linewidth]{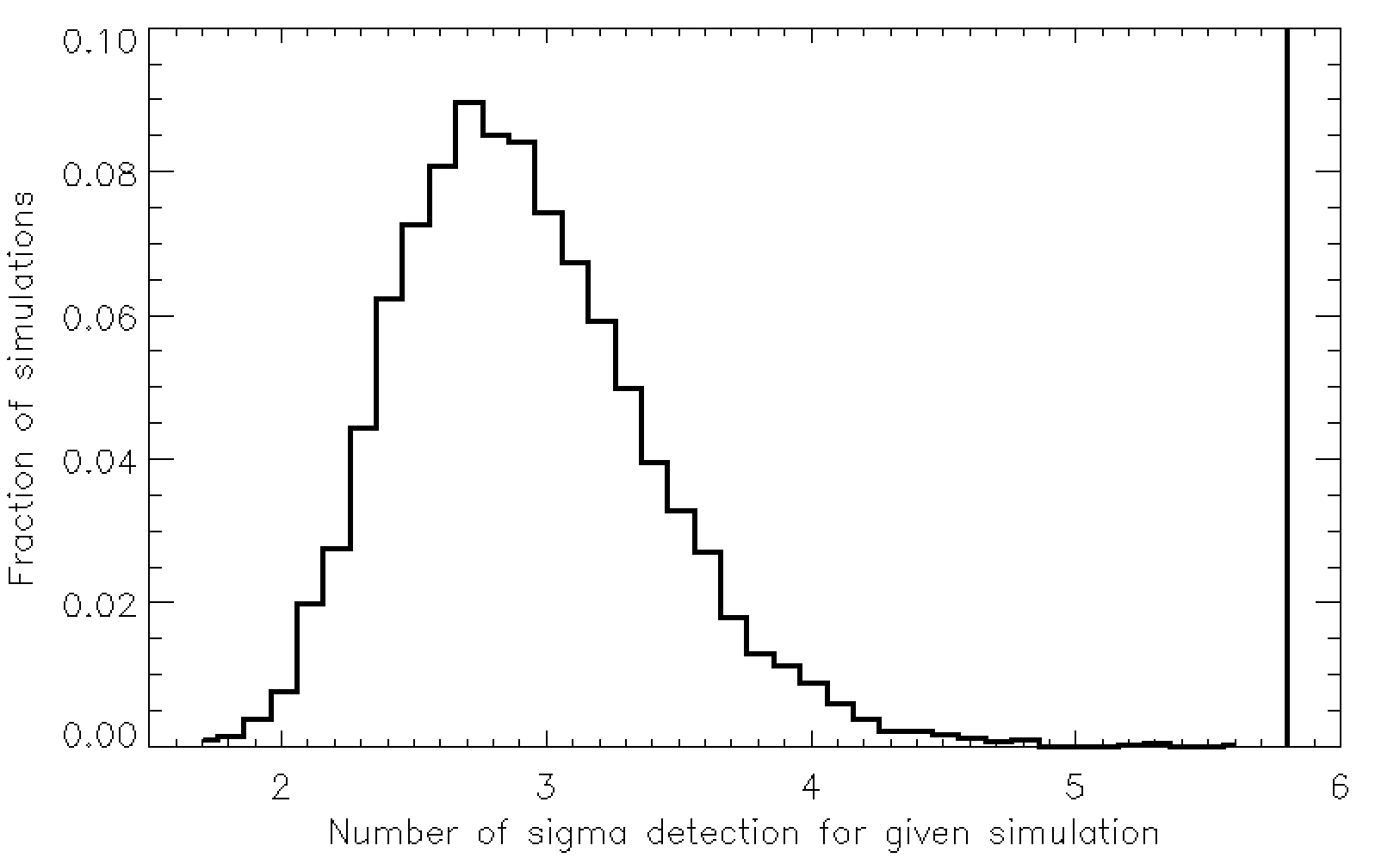}
  \caption{Same as Figure \ref{fig:devhist}, but now in addition to the other parameters, also adjusting for the multipole cut ($\ell>1$, $\ell>2$, $\ell>5$ or $\ell>10$) which optimizes detection for each simulation.\label{fig:devhisttotal}} 
\end{figure}

Finally, in addition the above parameters, we also allow the minimum multipole to vary. For the main $\ell$-ranges used in this paper, $\ell>1$, $\ell>2$, $\ell>5$ or $\ell>10$ we have calculated all profile depths for all 10.000 SMICA simulations allowing for an extended analysis where, for each simulation, the profile depth of the combination of minimum galaxy size, maximum redshift, galaxy type, inside/outside area ABC {\it and} minimum $\ell$-multipole which maximize the detection in either positive or negative temperature is recorded and shown in the histogram of Figure \ref{fig:devhisttotal}. We see that even when correcting for the look-elsewhere-effect for all these CMB and galaxy parameters, the data still show a stronger detection than 10.000 simulated maps.

\begin{figure}[htbp]
  \includegraphics[width=\linewidth]{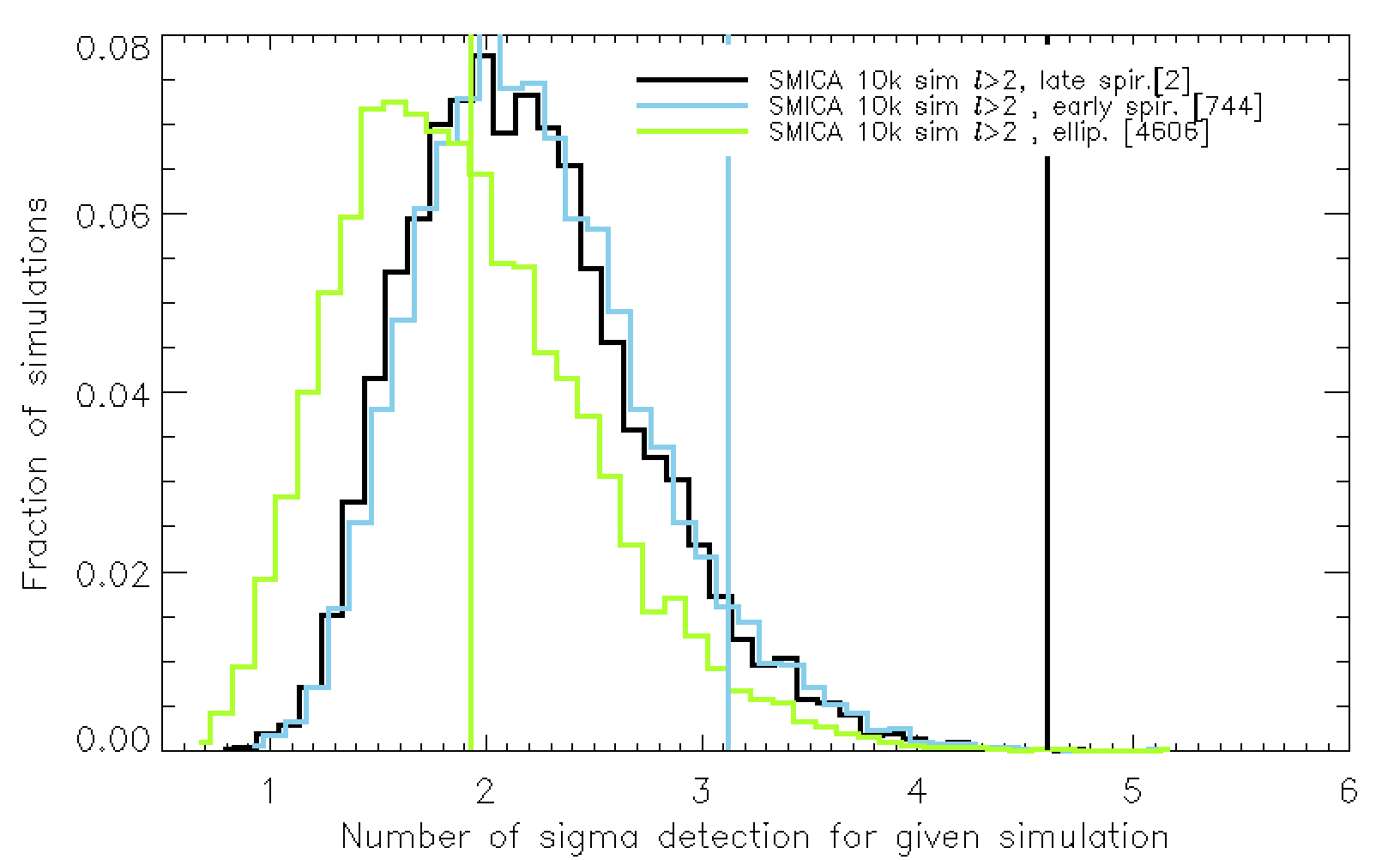}
  \caption{Same as Figure \ref{fig:devhist}, but only for the 10.000 SMICA simulations and with galaxy type fixed for the three different galaxy types, early spirals, late spirals and elliptical galaxies. The number in brackets show the number of simulations with a higher deviation than the data.\label{fig:devhisttypes}} 
\end{figure}

\subsection{Frequency dependence}

Finally, we look for frequency dependence of the profile depth, using the configuration with highest signal-to-noise and highest significance. For large spiral galaxies within areas A+B+C, we find for maps with $\ell\leq5$ removed a $5.7\sigma$ detection compared to simulations and a profile depth of $-27\mu$K. In C2024, they found a consistent signal profile over the full frequency range from $44$\;GHz to $217$\;GHz using the SEVEM frequency cleaned maps from the Planck release. Here we repeat the estimation of the profile depth for the same maps, using only areas A+B+C and with $\ell\leq5$ removed. The variations around $-27\mu$K are small and always $<1\mu$K. Comparing to the simulated SEVEM frequency maps, this tiny difference is significantly larger than found in simulations, but we found this difference consistent with expected thermal SZ effect in this area. Using the Planck derived map of the Compton $y$-parameter from \cite{sz}, we calculated an averaged $y$-parameter around the same galaxies used to estimate the profile depth. As these are spiral galaxies, no significant SZ effect is expected. We found an averaged $y$-parameter of $y=1.8\times10^{-7}$ which corresponds to the $\sim1\mu$K differences found between low and high frequencies.

Note that the SEVEM frequency cleaned maps are created using templates based on difference maps between other frequency channels. These templates are fitted and subtracted from a given frequency channel. If the signal is present in all frequency channels and also frequency dependent, the signal would also be present in the templates and the effect could be removed or amplified through the SEVEM template subtraction process. In order to avoid this possible under- or over-subtraction in the frequency maps, we tested the cleaned WMAP frequency maps. The cleaned publicly available WMAP maps at $41$\;GHz,  $61$\;GHz and $94$\;GHz for the final 9-year results \citep{wmap9} are also based on a template subtraction method with a combination of external templates, but also one internal template based on the difference between two frequency maps. This could again lead to under- or over-subtraction of the new foreground. However, in the WMAP 1st year results, only external templates were used for foreground subtraction. In this case, we do not expect the unknown foreground to be significantly affected by the template subtraction. We therefore apply the WMAP 1st year foreground subtraction method \citep{wmap1} on the 9 year data and still find no significant frequency dependence on the profile and the profile depth is consistent with the Planck profile depth. Also when the official foreground cleaned WMAP 9 year maps were used, no frequency dependence on the profile depth was found in the range $41$ - $94$\;GHz. We conclude that the unknown foreground is very stable over the full frequency range from $41$ - $217$\;GHz, consistent between the two experiments and different foreground subtraction methods.

\subsection{The role of elliptical galaxies}

In L2023, we found that the foreground signal mainly originates around late type spiral galaxies. It was shown that the inner parts of the temperature profile around elliptical galaxies show no significant signal, but with a foreground signal appearing and increasing at larger distances from the galaxy. As galaxies tend to cluster, this was interpreted as a signal from neighbouring spiral galaxies and not from the elliptical galaxies themselves. In C2024, temperature profiles for different types of galaxies including late spirals and elliptical galaxies, are shown to be very similar (their Figure 3B). They conclude that all types of galaxies, including elliptical galaxies, show a foreground signal, in contradiction to the results and conclusions of L2023.

Figure \ref{fig:devhisttypes} is similar to Figure \ref{fig:devhist}, but now only showing the result for the 10.000 SMICA simulations for the three galaxy types, early spirals, late spirals and elliptical galaxies. We clearly see that the signal is much more significant for late spiral galaxies and a possible $7\%$ effect for early type spiral galaxies. Elliptical galaxies however, do not show any foreground signal.\\

\begin{figure}[htbp]
  \includegraphics[width=\linewidth]{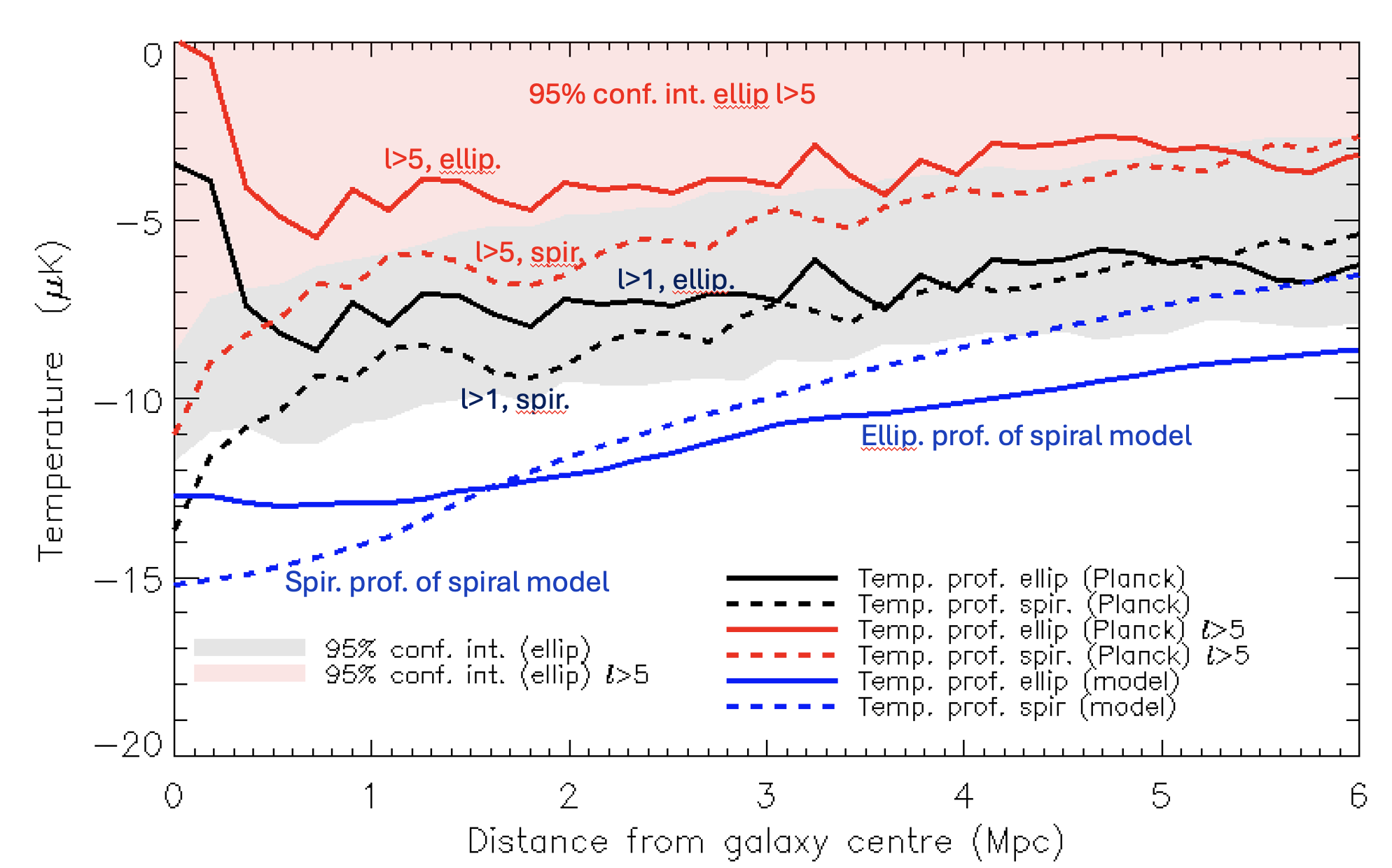}
  \includegraphics[width=\linewidth]{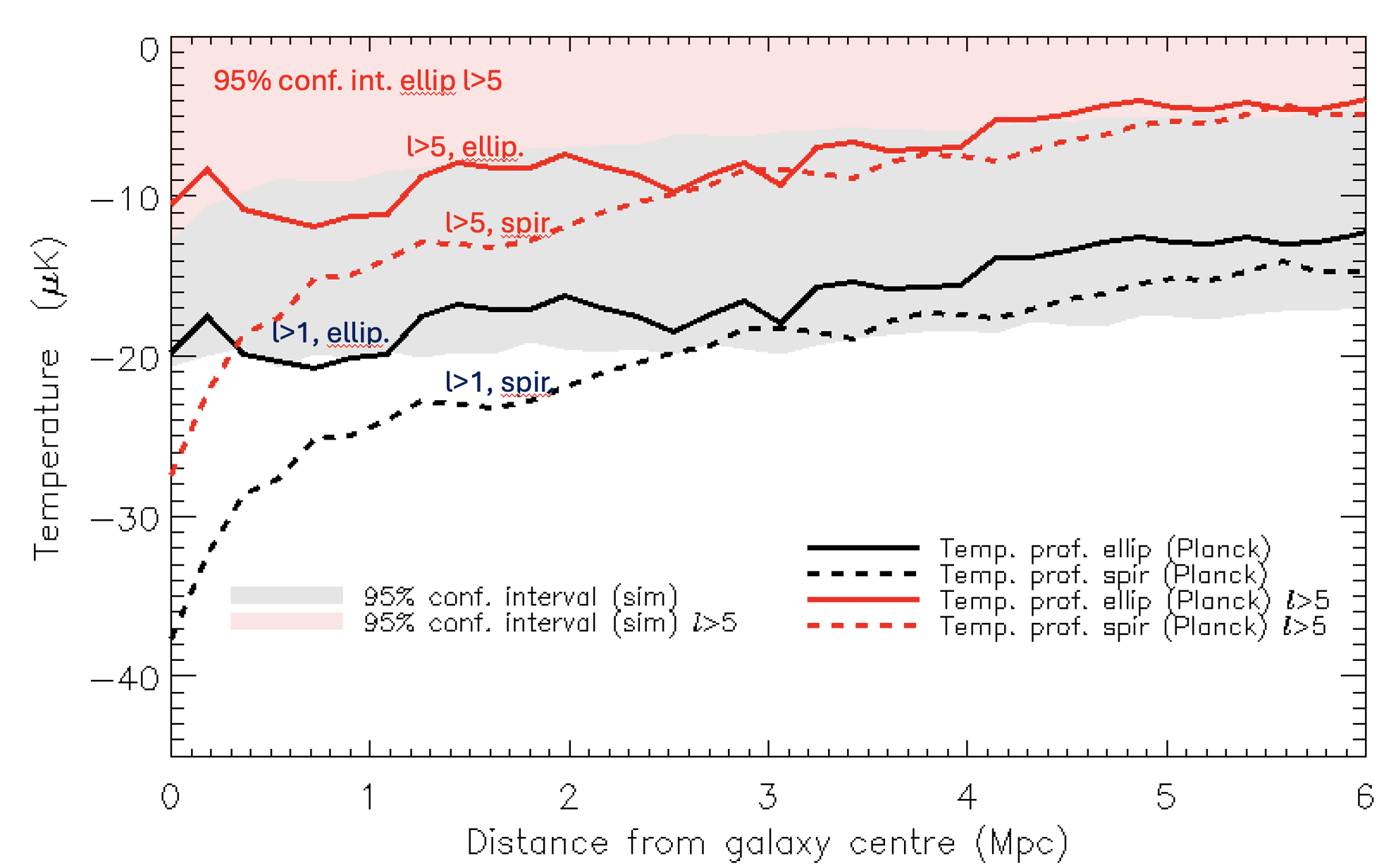}
\caption{Same plots as in Figure \ref{fig:longprof}, but now for elliptical galaxies (shaded areas and solid lines). The results for late type spiral galaxies from Figure \ref{fig:longprof} are present as dashed lines. Upper plot is the profile taken over galaxies in the full map, lower plot for galaxies within the 3 regions A, B and C only. The blue lines in the upper figure show the profiles calculated on the foreground model map of H2023 (shown in Figure \ref{fig:local}) where only late type spiral galaxies are assigned a synthetic profile. The solid blue line shows the profile taken around elliptical galaxies in this model whereas the dashed blue line shows the profile taken around late type spiral galaxies. \label{fig:longprof2}} 
\end{figure}

It is important to note that the number of elliptical galaxies is much smaller than the number of spiral galaxies. If the temperature profile for elliptical galaxies are indeed similar to the profile for spiral galaxies as claimed in C2024, the profile of the elliptical galaxies could still not show a significant foreground signal simply because the errors and thereby the distribution of profiles in simulations is much larger. In Figure \ref{fig:longprof2}, we show the profiles of SEVEM PR4 maps around elliptical galaxies out to 6 Mpc as we did for late spirals in Figure \ref{fig:longprof}. We also plot the profiles for late spirals as dashed lines for comparison. First of all, we can see, as expected due to the small number of ellipticals, that the distribution in the simulation is much wider comparing to the distribution for spirals in Figure \ref{fig:longprof}. For some of the bins, spirals and ellipticals show similar profiles in terms of temperature values, but for the elliptical galaxies these same values are not significant and mostly within the $95\%$ confidence level.

The upper plot of Figure \ref{fig:longprof2}, showing the profiles of galaxies in all environments, confirms the results of L2023. In this panel, the innermost 1-2 profile bins have nearly zero temperature, but for bins more distant than $0.5$\;Mpc, the signal is much stronger and to a large degree follows the profile of spiral galaxies. In the lower plot, showing galaxies in high density environments, the change in bin 1-2 is not seen and the profile is consistently low over larger distances, but again not significant due to the large uncertainty.

In C2024, the inner $0.2^\circ$ of the profile is not shown. In that paper, the first profile bin starts at an angular distance of $0.2^\circ$ from the galaxy and can therefore be considered consistent with our results since the part of the profile which shows almost zero temperature is omitted in their work. We therefore find full consistency with the profiles of C2024 taking into account the same part of the profile (although a direct comparison is difficult since they use angular distances and not projected distances in Mpc as we do here). The reason for masking the inner $0.2^\circ$ as was done in C2024 but also in H2023, is to avoid possible emission from the visible galaxy. Since such an emission would necessarily {\it increase} the temperature profile and in this way lower the detection of the new foreground. We therefore consider the results using this inner bin a conservative result. Still we have tested using the profile bins within $[0.2\textrm{Mpc},0.6\textrm{Mpc}]$ instead of the profile depth and find that the results reported here are not significantly altered.

In L2023 we suggested that the profile for the ellipticals shown in the upper plot of Figure \ref{fig:longprof2} where there is almost zero average temperature close to the elliptical galaxies can be explained by neighbouring galaxies giving the signal at larger distances. This is also consistent with the fact that the profile in the lower plot of Figure \ref{fig:longprof2} for denser environments, shows low temperatures also closer to the elliptical galaxies. In the upper plot of Figure \ref{fig:longprof2}, we show in blue colour the profiles for spirals and ellipticals calculated on the foreground model map of H2023 (shown in Figure \ref{fig:local}). This map was created assigning synthetic profiles with varying amplitude and widths depending on galaxy size and environment to the late spiral galaxies of 2MRS. We see that although a profile was only assigned to late type spiral galaxies, the temperature profile around spiral and elliptical galaxies are similar, as we also see in the actual data. We cannot rule out the possibility that elliptical galaxies also show a foreground signal. If ellipticals had emission from the visible part of the galaxy cancelling the new foreground signal and this emission is not seen in spirals, the profiles of Figure \ref{fig:longprof2} could be explained. However, the above evidence and arguments seem to indicate a much weaker or non-existent foreground signal from elliptical galaxies. This is also supported by the observation in GL2024 that the nearby elliptical dominated Fornax cluster does not show any low temperature feature in the CMB whereas the foreground signal from its neighbouring spiral dominated Eridanus group may be responsible for the CMB Cold Spot, as argued in GL2024.

\subsection{Low multipole correlations}

\begin{figure*}[htbp]
\includegraphics[width=0.5\linewidth]{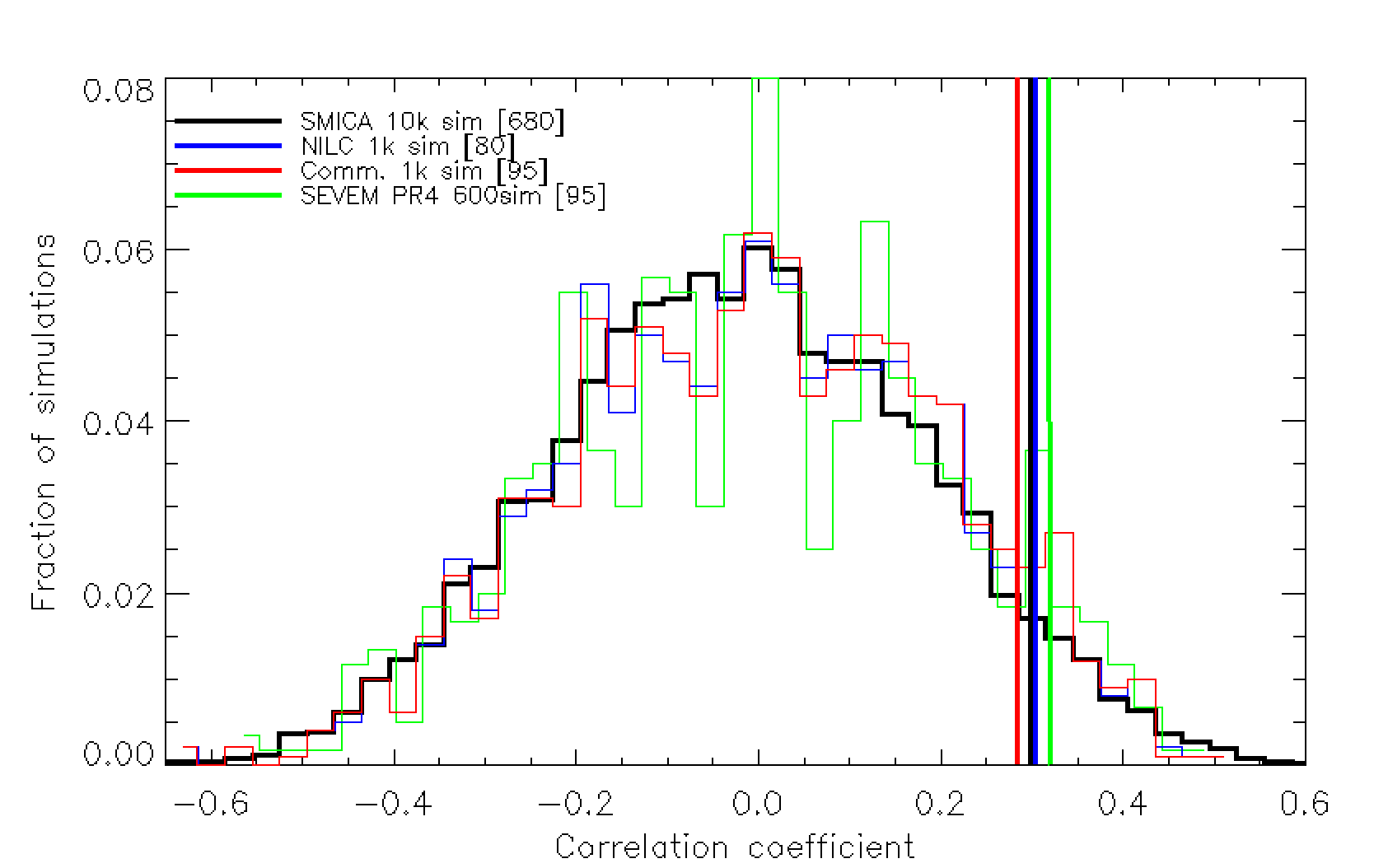}
\includegraphics[width=0.5\linewidth]{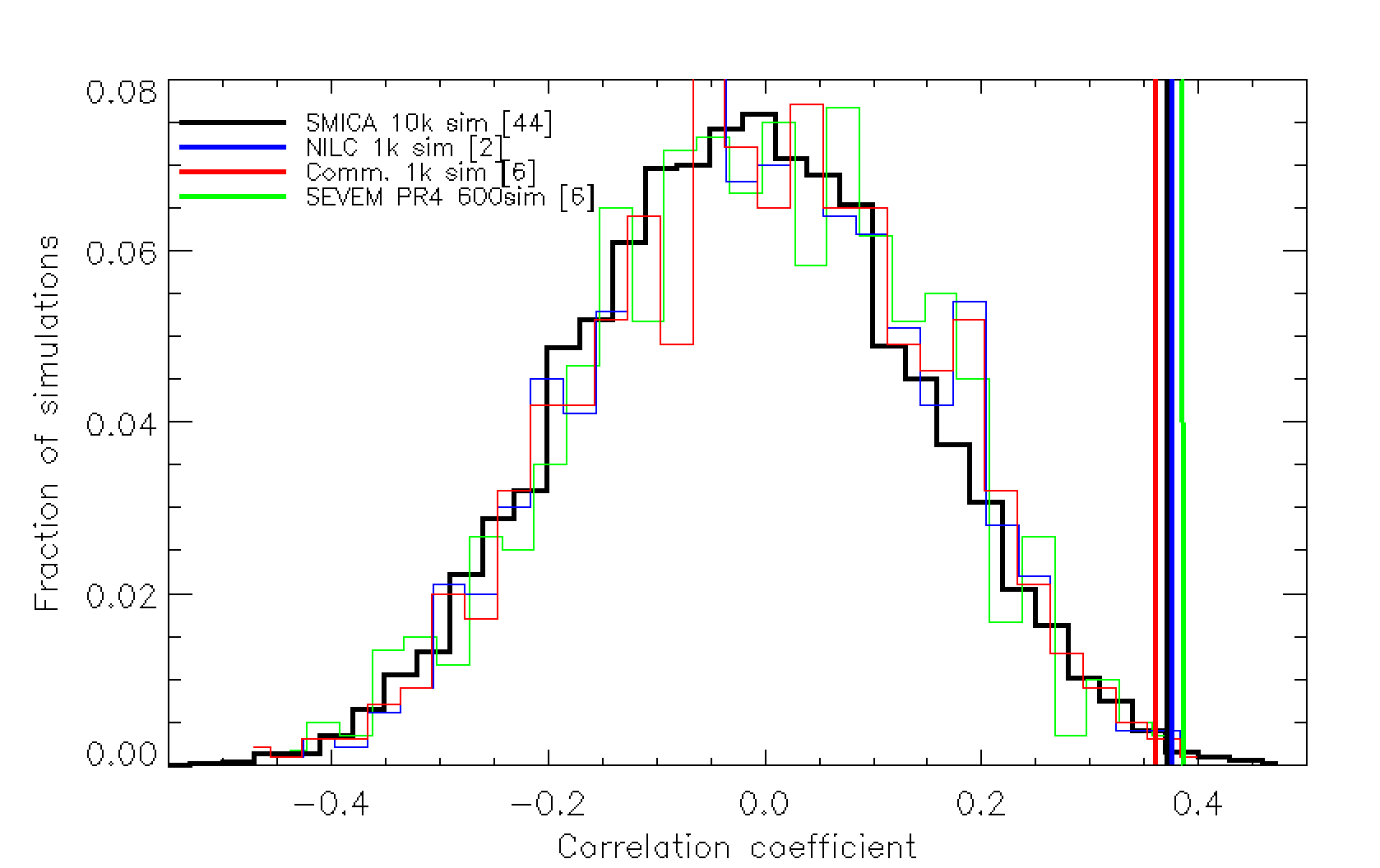}
\includegraphics[width=0.5\linewidth]{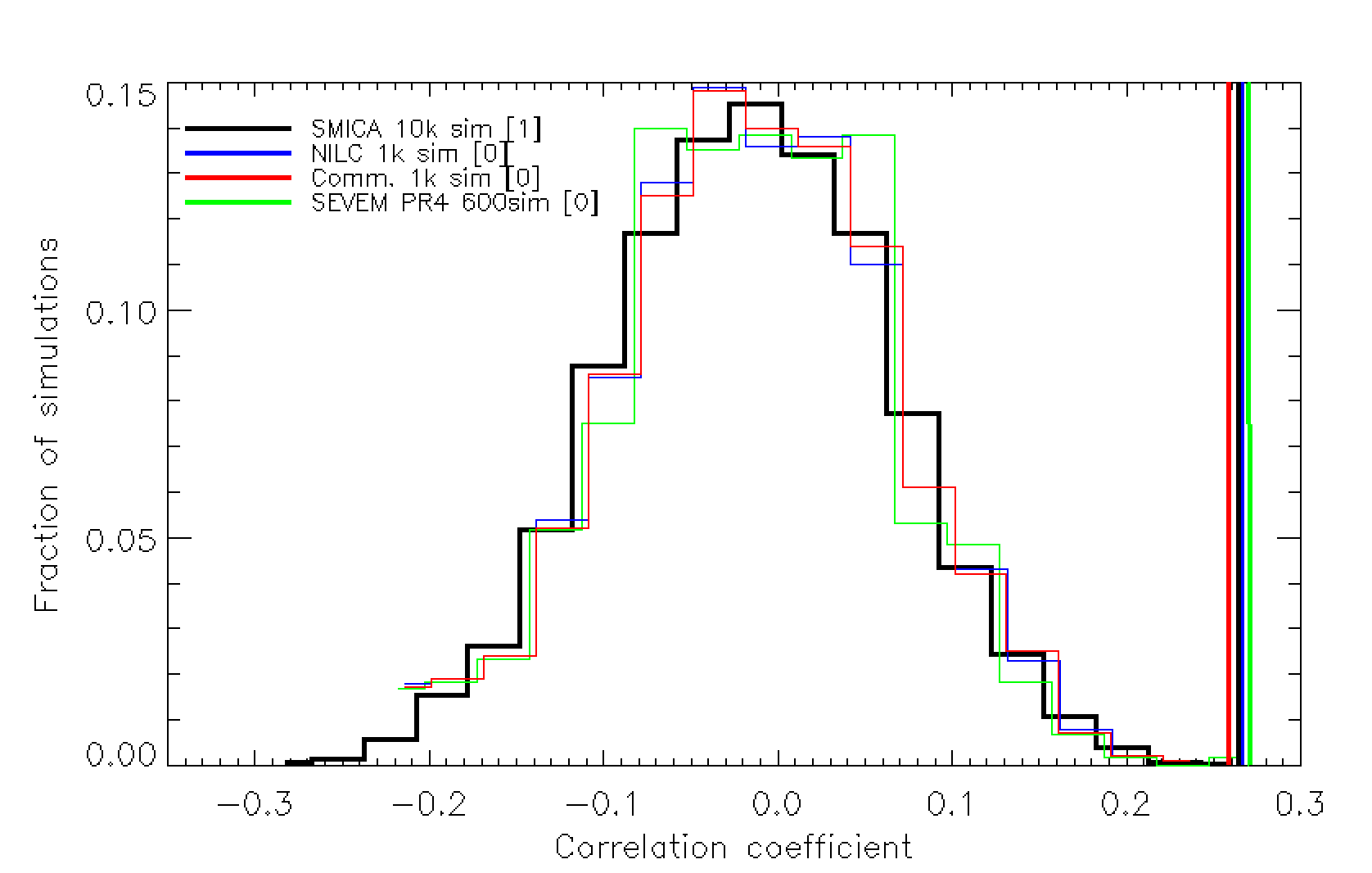}
\includegraphics[width=0.5\linewidth]{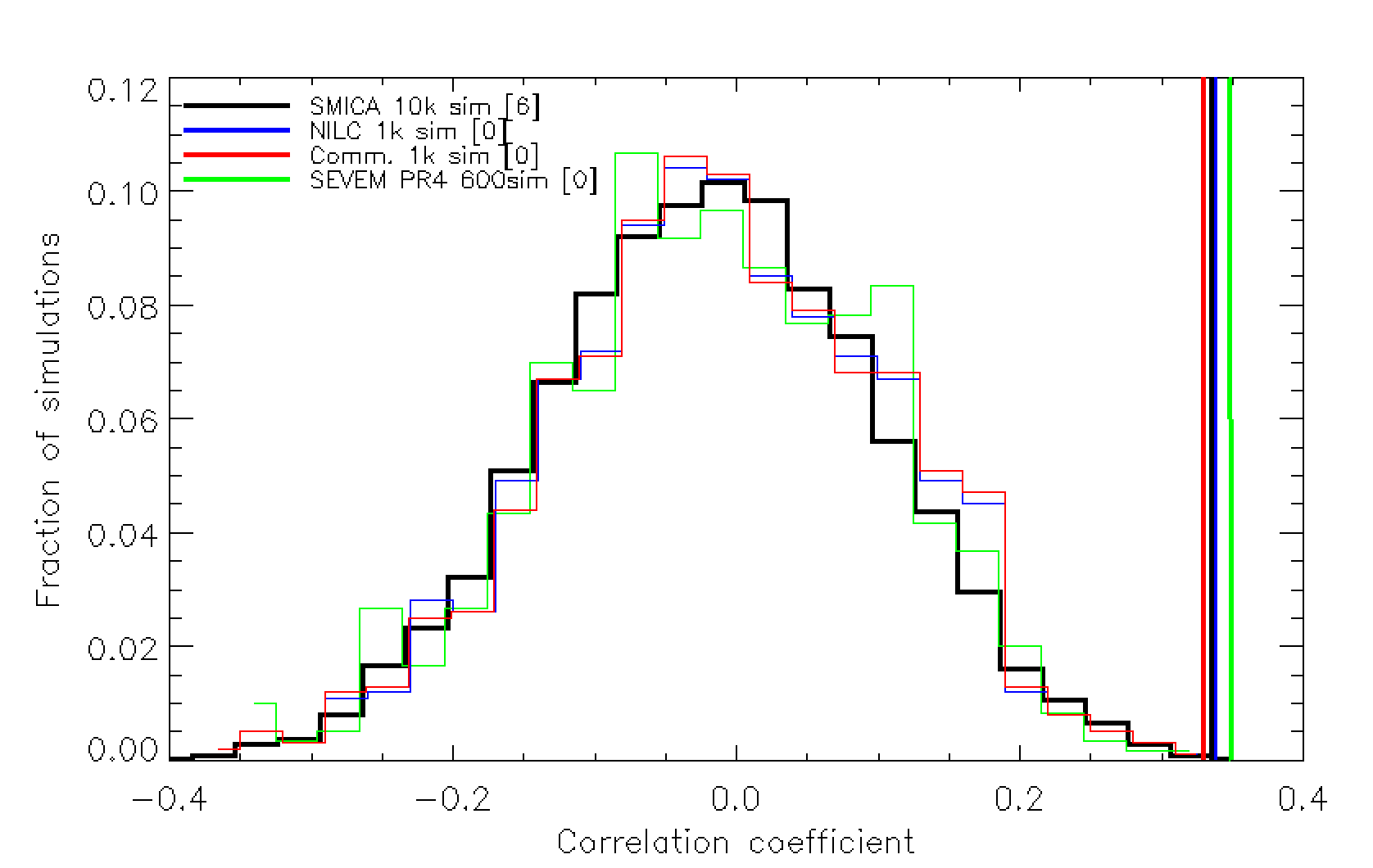}
\caption{Correlation coefficients between CMB maps and the galaxy distribution map for low multipoles. Histogram shows correlation coefficients between the galaxy distribution map and simulated CMB maps, vertical line shows the same using the Planck CMB map. Clockwise from the upper left plot we show the correlation results with upper multipoles $\ell_\mathrm{max}=5$, $\ell_\mathrm{max}=7$, $\ell_\mathrm{max}=10$ and $\ell_\mathrm{max}=16$. The number in brackets show the number of simulations with a higher correlation coefficient than the data.\label{fig:corrhist}} 
\end{figure*}

Finally, Figure \ref{fig:corrhist} shows the distribution of correlation coefficients between the normalized low multipole maps of the CMB simulations and galaxy model map. The plots show results when including all multipoles less than or equal to $\ell_\mathrm{max}=5$, $\ell_\mathrm{max}=7$, $\ell_\mathrm{max}=10$ and $\ell_\mathrm{max}=16$ for the maps reduced to {\texttt HEALPix} resolution $N_\mathrm{side}=4$. Consistent with the above results for the profiles, the largest scales $\ell\leq5$, show little correlation. For $\ell>5$ the correlation increases rapidly as one would expect given the angular extent of the foreground signal. For $\ell_\mathrm{max}=16$, only 1 of 10.000 CMB simulations shows a larger correlation with the galaxy model map than the Planck data itself.

We have thus confirmed the presence of the signal for large angular scales and small angular scales independently. For the largest angular scales, we find a strong correlation with the galaxy distribution map. When removing these scales from the CMB map, we still find a very significant negative profile depth around these same galaxies in the smaller scales.

\section{Discussion and Conclusion}
\label{sect:concl}

We have confirmed that the average CMB temperature in discs around large spiral galaxies in average is lower compared to the corresponding mean disc temperature measured in simulated CMB maps at the $3\sigma$ level. Separating the largest and smallest scales of the CMB maps, we find strong correlations with the galaxy distribution independently in the two multipole ranges. Removing the largest scales increases signal-to-noise for the average disc temperature and increases the significance to the $5.7\sigma$ level for galaxies in the regions of the nearby cosmic filaments denoted by A, B and C in Figure \ref{fig:local}. Of 10.000 simulated maps, none show a similarly low average temperature around galaxy positions. For the largest scales, we find that the correlation between the low multipoles $\ell<16$ of the Planck CMB map and a map of nearby large spiral galaxies is significant at the $99.99\%$ level. Correcting for the look-elsewhere-effect, we find the result to be very stable with respect to choices of galaxy samples with different parameters such as redshift and size.

With such a strong detection, the existence of an unknown process removing or cooling CMB photons in and around galaxies has been confirmed. In A2024, the significance of the earlier detections was disputed, but here we have corrected some common weaknesses in the approaches of both A2024 and previous papers (L2023, H2023), showing now that the significance is much larger. Instead of a $\chi^2$ approach including temperatures at huge distances from the galaxies, we focus our attention to a disc around the galactic centre where the signal is expected to be found if the physical origin is related to the galactic halo.

The physical mechanism behind the signal is hitherto unknown. Given the foreground signal extends far beyond the extension of the visible galaxy, an explanation in terms of a dark matter induced signal is speculative but highly relevant. If the signal arises from interactions of photons with dark matter particles in galactic halos, the process needs to preserve the black body spectrum over Planck frequencies. We discuss above that the temperature profiles observed around elliptical galaxies may be interpreted as a signal from the neighbouring spiral galaxies and not originating in the elliptical galaxies themselves. If this were the case, a hypothesis of photons interacting with dark matter appears inconsistent. The main differences between spiral galaxies and elliptical galaxies is the rotation and stronger magnetic fields of the former. Another difference is the higher density of ionized gas in elliptical galaxies giving rise to Sunyaev-Zeldovich effect and X-ray radiation from clusters of galaxies.

One step on the path to an explanation is to find the properties of the galaxies causing the signal and in particular to test of it is the same kind of galaxies giving rise to the signal in different parts of the sky. By repeating the above procedure, searching for the galaxy properties which optimize the foreground signal in each of the regions A, B and C separately (see Figure \ref{fig:local}), we find that in all three regions, the late type spiral galaxies are the ones giving the lowest profile depth. In Table \ref{tab:abc} we show the redshift and size ranges (allowing both lower and upper redshift and size ranges to vary) which give the largest detection in each of these areas. The redshift ranges are very consistent suggesting that in general the signal originates from late type spiral galaxies with size larger than about $r=8$ kpc and for redshifts less than about $z=0.02$. The reason for this upper limit in redshift may arise from the increased uncertainty in galaxy morphology determination as discussed above.

The galaxies in the three regions A, B and C that alone appear to give rise to the entire foreground signal, divide the total sample of nearby large spiral galaxies into two approximately equal fractions. The galaxies within these regions reside in the most dense galactic environments whereas the mean density around the galaxies outside of these regions is much smaller. These areas also contain nearby dense cosmic filaments and comprise the most massive spiral dominated galaxy groups within $z<0.02$. The exact position and radius of these three circles were chosen by-eye, but we have seen that the results do not change substantially with small shifts in position and radius as long as the densely populated filaments are kept inside the rings. For instance increasing or decreasing the radius of these areas with $5^\circ$, the significance remains $>5\sigma$. The foreground signal clearly originates in environments with a high density of large spiral galaxies. We cannot exclude that the signal is also present in less dense areas, a weaker signal could be confused by the CMB fluctuations themselves.

We have tested whether the signal mainly originates in the galaxy position centres in the groups within these filaments or otherwise, if also galaxies outside of the groups contribute significantly. Looking at spiral galaxies within 2 Mpc of the galaxies in these groups (adopting a minimum group mass of $10^{11}\textrm{M}_\odot$), we find a $5\sigma$ detection alone for these about 700 galaxies. For the remaining 500 galaxies which are in projection to these dense cosmic filaments (areas A, B and C), but outside the large groups in these areas, we still find a $3\sigma$ detection. The signal is stronger close to the spiral dominated galaxy groups, but it is clearly present also in less dense environments within the dense filaments. We also find that displacing the galaxy positions randomly by fractions of a Mpc, the significance starts declining, showing that the foreground signal is indeed related to the galaxy positions, although some part of the signal persists to several Mpc outside the galaxies.

Given that the foreground signal appears concentrated in dense galactic environment with large spiral galaxies as a tracer, the hypothesis of a relation to magnetic fields is highly plausible. CMB photons arriving from the parts of the sky where the new foreground is strongly present have travelled through several Mpc of galactic and intergalactic magnetic fields. Some hypothesised dark matter particles, in particular Axion Like Particles (ALP), are known to have the possible property of being created from photons in magnetic fields, and given the time the photons spend passing through the galactic fields in dense galactic environments, the probability of this to occur would be much higher for these CMB photons (See \cite{axionlss} and references therein for a study of this effect on large scale structure). Intriguing results on the existence of ALPs have been found from measurements of cosmic birefringence, see \cite{biref} for a recent update.

The effect of photons being converted to ALPs has been studied in the context of clusters of galaxies \citep{axioncluster1,axioncluster2}, but as shown in these works, spectral distortions are expected, inconsistent with our observations. The frequency dependence may depend on properties of the magnetic field and the plasma through which the CMB photons pass \citep{axionbfield,axionlss}, but we find that the observed temperature decrement only changes by at most a few percent between $41$\;GHz and $217$\;GHz. If the hypothesis of photons being converted to unknown particles in magnetic fields is correct, one would have to look for particles for which this process to a large degree maintains the CMB black body shape in the microwave range of the electromagnetic spectrum. The strong galaxy density dependence of the effect makes hypothesized particles which change properties depending on the environment, chameleon particles, another possible candidate (see i.e. \cite{cham1,cham2}).

\begin{table}[htbp]
  \begin{center}
    \caption{The best fit z- and size range of late spiral galaxies within each of the areas A, B and C.}
    \label{tab:abc}
    \begin{tabular}{c|c|c}
      \textbf{galaxies in...} & \textbf{ z-range} & \textbf{size range} \\
\hline
Area A & $z=[0.004,0.019]$ & [$7$\; kpc, $15$\; kpc] \\
Area B & $z=[0.004,0.020]$ & [$9$\; kpc, $16$\; kpc] \\
Area C & $z=[0.004,0.016]$ & [$9$\; kpc, $23$\; kpc] \\
\hline
    \end{tabular}
  \end{center}
  \tablefoot{The z-range and size range of late spiral galaxies within each of the areas A, B and C which gives the most significant profile depth, requiring at least 200 galaxies in the sample. Both lower and upper redshift and size ranges were adjusted in order to optimize profile depth. Results based on SMICA maps.}
\end{table}

An important next step for understanding the origin of the cooling of the CMB around galaxies is to test the presence of a similar effect at larger redshifts $z>0.02$. In a forthcoming study, we will use other redshift catalogues with more reliable morphological classifications of the galaxies at higher redshift. Although there are at present no other full-sky redshift catalogues available, increasing the redshift ranges in catalogues with a limited sky coverage may still maintains a high number of galaxies and thereby low uncertainties. The CMB polarisation signal is another important source of information which could shed light on the origin of the new foreground and which will be studied in a forthcoming paper.

\begin{acknowledgements}
We are thankful to Sigurd K. Næss for fruitful discussions and suggestions. Results in this paper are based on observations obtained with Planck (http://www.esa.int/Planck), an ESA science mission with instruments and contributions directly funded by ESA Member States, NASA, and Canada. We acknowledge the use of NASA’s WMAP data from the Legacy Archive for Microwave Background Data Analysis (LAMBDA), part of the High Energy Astrophysics Science Archive Center (HEASARC). The simulations were performed on resources provided by UNINETT Sigma2 - the National Infrastructure for High Performance Computing and  Data Storage in Norway". Some of the results in this paper have been derived using the HEALPix package \citep{healpix}
\end{acknowledgements}

\end{document}